\documentclass[twocolumn]{aastex63}
\usepackage{color,comment}
\usepackage[normalem]{ulem}
\usepackage[caption=false]{subfig}
\usepackage{graphicx}
\usepackage{verbatim}
\usepackage{mathtools}

\usepackage{float}
\usepackage{wrapfig}

\newcommand{\Msun}[1]{M$_{\odot}$}

\received{}
\revised{}
\accepted{}

\shorttitle{Dust Attenuation in Galaxy SED Fitting}
\shortauthors{S. Lower et al.}

\begin{document}

\title{How Well Can We Measure Galaxy Dust Attenuation Curves? \\ The Impact of the Assumed Star-Dust Geometry Model in SED Fitting}

\author[0000-0003-4422-8595]{Sidney Lower}
\affil{Department of Astronomy, University of Florida, 211 Bryant Space Science Center, Gainesville, FL, 32611, USA}
\author[0000-0002-7064-4309]{Desika Narayanan}
\affil{Department of Astronomy, University of Florida, 211 Bryant Space Science Center, Gainesville, FL, 32611, USA}
\affil{University of Florida Informatics Institute, 432 Newell Drive, CISE Bldg E251 Gainesville, FL, 32611, US}
\affil{Cosmic Dawn Centre at the Niels Bohr Institue, University of Copenhagen and DTU-Space, Technical University of Denmark}
\author[0000-0001-6755-1315]{Joel Leja}
\affil{Department of Astronomy \& Astrophysics, The Pennsylvania State University, University Park, PA 16802, USA}
\affil{Institute for Computational \& Data Sciences, The Pennsylvania State University, University Park, PA, USA}
\affil{Institute for Gravitation and the Cosmos, The Pennsylvania State University, University Park, PA 16802, USA}
\author[0000-0002-9280-7594]{Benjamin D. Johnson}
\affil{Center for Astrophysics | Harvard \& Smithsonian, 60 Garden St. Cambridge, MA 02138, USA}
\author[0000-0002-1590-8551]{Charlie Conroy}
\affil{Center for Astrophysics | Harvard \& Smithsonian, 60 Garden St. Cambridge, MA 02138, USA}
\author[0000-0003-2842-9434]{Romeel Dav{\'{e}}}
\affil{Institute for Astronomy, Royal Observatory, University of Edinburgh, Edinburgh, EH9 3HJ, UK}
\affil{Department of Physics and Astronomy, University of the Western Cape, Bellville, 7535, South Africa}
\affil{South African Astronomical Observatories, Observatory, Cape Town 7925, South Africa}

\begin{abstract}

One of the most common methods for inferring galaxy attenuation curves is via spectral energy distribution (SED) modeling, where the dust attenuation properties are modeled simultaneously with other galaxy physical properties. In this paper, we assess the ability of SED modeling to infer these dust attenuation curves from broadband photometry, and suggest a new flexible model that greatly improves the accuracy of attenuation curve derivations. To do this, we fit mock SEDs generated from the {\sc simba} cosmological simulation with the {\sc prospector} SED fitting code. We consider the impact of the commonly-assumed uniform screen model and introduce a new non-uniform screen model parameterized by the fraction of unobscured stellar light. This non-uniform screen model allows for a non-zero fraction of stellar light to remain unattenuated, resulting in a more flexible attenuation curve shape by decoupling the shape of the UV attenuation curve from the optical attenuation curve. The ability to constrain the dust attenuation curve is significantly improved with the use of a non-uniform screen model, with the median offset in UV attenuation decreasing from $-0.30$ dex with a uniform screen model to $-0.17$ dex with the non-uniform screen model. With this increase in dust attenuation modeling accuracy, we also improve the star formation rates (SFRs) inferred with the non-uniform screen model, decreasing the SFR offset on average by $0.12$ dex. We discuss the efficacy of this new model, focusing on caveats with modeling star-dust geometries and the constraining power of available SED observations.
\end{abstract}

\section{Introduction}

Dust, though a subdominant portion of a galaxy's mass, has a significant impact on galaxy spectra: up to half of stellar light integrated from $z = 0 - 8$ has been obscured and reprocessed by dust \citep{madau_dickinson_2014, casey_2014, casey_2018, zavala_2021}. Understanding the underlying stellar population and gas content in a galaxy, including stellar and gas masses, star formation rates, and metallicity, depends on simultaneously modeling the dust absorption and emission of a galaxy to uncover the intrinsic stellar and nebular emission \citep{walcher_2011_araa, conroy_2013_araa}. This is done by applying a dust attenuation curve, which defines the amount of attenuation as a function of wavelength, and is a function of dust properties and the observed distribution of stars and dust in the galaxy.

An important distinction to make is the difference between a dust \textit{extinction} curve and a dust \textit{attenuation} curve. Dust extinction is dependent on the properties of dust (i.e., the grain size distribution and composition), and is measured along a single sight line towards a back-lit region. Dust attenuation folds in the effects arising from the distribution of stars and dust in the galaxy, such that attenuation accounts for the effective amount of light lost in aggregate for a number of sight-lines that includes both the contribution of scattering back into the line of sight, as well as contributions from unobscured stars. Thus, while extinction measurements are limited to nearby objects, where individual lines-of-sight can be resolved, attenuation can potentially be measured for objects out to high redshifts \citep[for a recent review, see][]{attenuation_araa}.

However, measuring the dust attenuation of stellar continuum spectra is difficult to achieve in practice. The pioneering work of \citet{calzetti_1994},  \citet{calzetti_1997} and \citet{calzetti_2000} measured the attenuation of local starburst galaxies using Balmer decrements, which measure the attenuation of \ion{H}{2} regions surrounding massive stars. The attenuation curve of \citet{calzetti_2000} was fit as a polynomial function of 1/$\lambda$ and features a shallower slope compared to the average Milky Way extinction curve and does not include the 2175 $\AA$ bump found in several other curve models and parameterizations. 

The general features of attenuation curves include the shape, typically parametrized by the ratio of attenuation at 1500 $\AA$ to the attenuation in the V-band, the existence of a bump feature at 2175 $\AA$, and the absolute attenuation in the optical V-band. In the local Universe and at intermediate redshifts, studies have demonstrated a wide diversity in galaxy dust attenuation curves. Correlations have been found between attenuation curve shape and galaxy physical properties like stellar mass and star formation activity \citep[e.g.][]{noll_2007, noll_2009, salim_2016, battisti_2016, battisti_2017_1, corre_2018, salim_2018, reddy_2018, cleri_2020_pabeta, Bogdanoska_2020}, galaxy spectral type \citep[e.g.][]{kc_law}, morphology and observed inclination angle \citep[e.g.][]{battisti_2017_2, zuckerman_2021_inclination}, and the optical properties of the dust grains \citep[e.g.][]{tress_2018, attenuation_araa}.

One of the most common methods for inferring the shape of galaxy attenuation curves in large galaxy samples is via spectral energy distribution (SED) modeling, where the dust attenuation properties are modeled simultaneously with other galaxy physical properties. Full SED modeling is necessary to account for each components' (i.e., stellar continuum, dust reddening, metallicity) influence on galaxy colors. Even then, degeneracies between model parameters make determining the true dust attenuation curve challenging \citep[e.g][]{2022MNRAS.tmp..190Q} as these degeneracies can influence inferred relations between attenuation curve parameters. Still, the most robust way to determine the attenuation properties of galaxies is via SED modeling. For instance, using  {\sc fsps} stellar population modeling \citep{fsps_1, fsps_2} and the {\sc FAST} fitting routine \citep{fast_software}, \citet{kc_law} found that the strength of the $2175 \AA$ bump is dependent on galaxy spectral type and the slope of the attenuation curve. \cite{salim_2018} fit galaxies at $z < 0.3$ from the GALEX-SDSS-WISE Legacy Catalog (GSWLC) with the SED modeling code {\sc cigale} \citep{cigale_1, cigale_2, boquien_2019_cigale} and found that a majority of galaxies have attenuation curves with significantly steeper slopes than the \cite{calzetti_2000} curve.  

Folded into the dust attenuation model are assumptions about the star-dust geometry in the galaxy. It is extremely difficult to infer the spatial distribution of the stars and dust from unresolved observations primarily because the impact of the star-dust geometry on a dust attenuation curve is partially degenerate with the optical properties of the dust \citep{witt_gordon_1996, witt_gordon_2000, hagen_2017, corre_2018}. Thus, as a simplifying approximation, the dust is assumed to be a uniform screen, in which all stars sit behind dust of constant column density and all stellar light is attenuated by dust with equal optical depth. To accommodate the impact of higher optical depths towards star forming regions, a variation of this model is that of \citet{charlot_fall_2000}, in which the dust attenuation is different for young ($< 10$ Myr) and old stellar populations, on the basis that younger stars are still coupled with their dense birth clouds and thus experience an additional source of dust attenuation. The UV luminosity and nebular emission lines associated with the young, massive stars are more obscured than the optical light dominated by the longer-lived stars. If nebular emission is considered in the SED fit, this additional source of attenuation for young stars is also applied to the emission lines from these same regions; typical average values for the ratio of reddening in nebular regions to reddening of stellar continuum range from $2-4$ \citep{calzetti_2000, reddy_2015, reddy_2020_neb, cleri_2020_pabeta}. 

The fundamental goal of this paper is to assess, given the aforementioned uncertainties in dust attenuation modeling, how accurately we can derive attenuation curves from traditional SED fitting.  We additionally introduce an alternative model to the standard uniform screen approximation that allows for a much wider, physically-motivated range of dust attenuation curves. We demonstrate that we are able to accurately and simultaneously infer the shape of galaxy attenuation curves and galaxy physical properties. We do this by generating mock SEDs for galaxies formed in a cosmological simulation, and fitting these broadband SEDs with three attenuation models with increasing complexity. We focus on the attenuation by diffuse dust only, neglecting the impact from higher density birth clouds and nebular regions and as such, all analysis focuses on the reddening of stellar continuum only. 

 \begin{figure*}
    \centering
    \includegraphics[width=\textwidth]{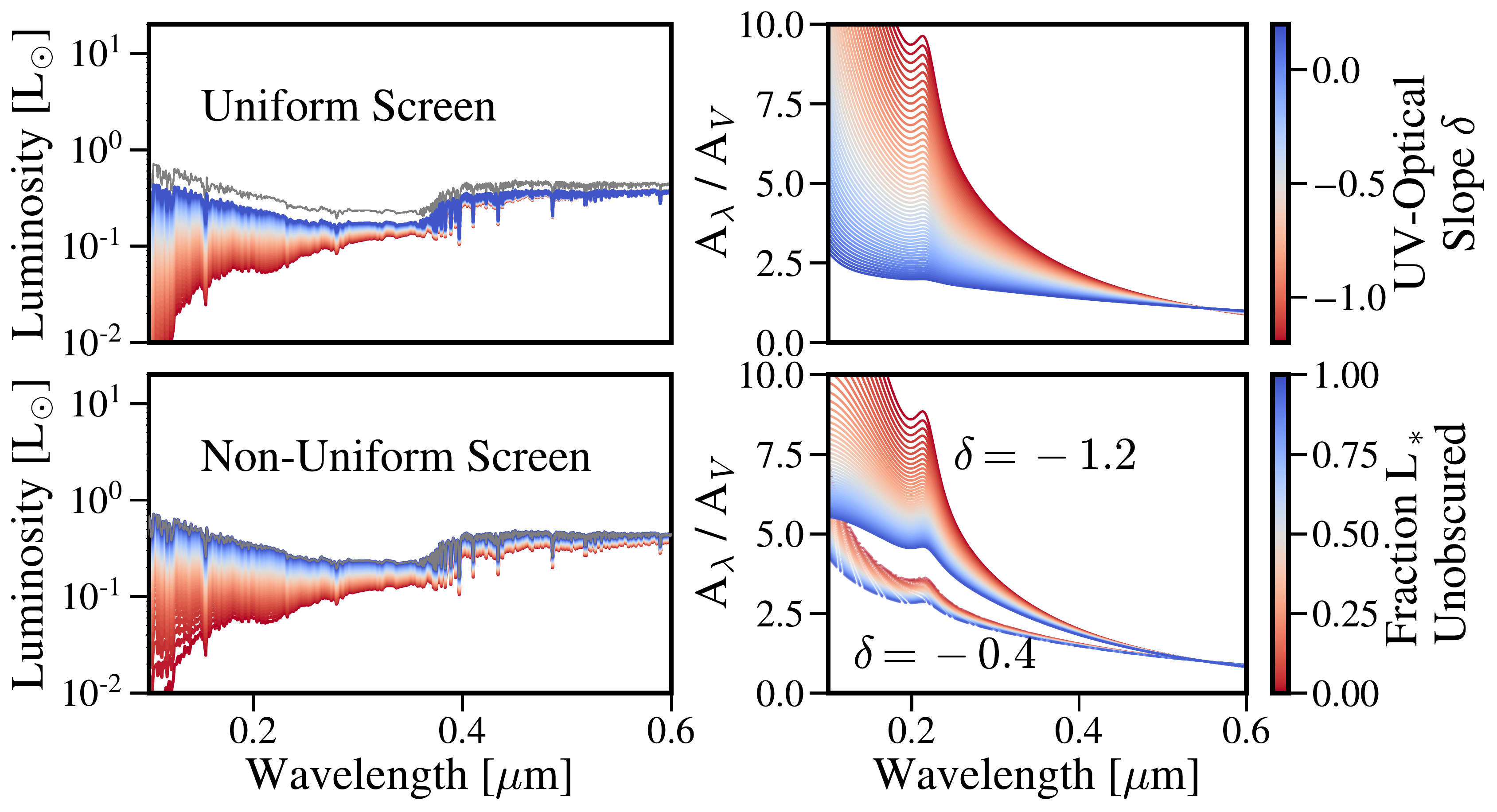}
    \caption{Comparison of SEDs and attenuation curves generated from (Top): a model with a variable UV-optical slope parameterized as the power-law deviation ($\delta$) from the \citep{calzetti_2000} curve. (Bottom): a model with a variable fraction of unattenuated stellar light. We show the spread in attenuation curves from increasing values of unobscured stellar light for two values of $\delta$. For a fixed value of $\delta$, varying f$_\mathrm{unobscured}$ results in attenuation curves decoupled from a power-law deviation of the Calzetti curve shape.}
    \label{fig:frac_vs_slope_shape}
\end{figure*}

The paper is outlined as follows: in Section \S\ref{sec:toy_models}, we discuss the primary drivers of attenuation curve variations and how these variations are traditionally modeled in SED fitting; in Section \S\ref{sec:methods}, we describe the {\sc simba} galaxy formation model, the post-processing {\sc powderday} 3D dust radiative transfer model, and the {\sc prospector} SED models; in Section \S\ref{sec:results}, we present the results of the {\sc prospector} SED fitting, focusing on the inferred attenuation curves and galaxy physical properties; and in Section \S\ref{sec:caveats} we discuss the caveats to our results, focusing on the issue of constraining a more flexible model in the context of Bayesian evidence and available photometry.

\section{Understanding Variations in Dust Attenuation Curves} \label{sec:toy_models}
 
The nature of a galaxy's dust attenuation depends on the dust extinction curve, which in turn depends on the grain size distribution, and optical properties of the grains, and the star-dust geometry, including dependencies on the stellar age distribution. Even with a fixed dust extinction curve, dust attenuation curves can vary due to the relative positions of the stars and dust \citep{calzetti_1994, calzetti_2001_geo,witt_gordon_1996,witt_gordon_2000}. For a fixed dust extinction curve, we expect galaxy attenuation curves to be diverse due to increasingly complex star-dust geometries that vary as a function of galaxy type and redshift as well as for varying inclination angles \citep{seon_draine_2016, popping_2017, narayanan_2018_atten, trayford_2020,  attenuation_araa, zuckerman_2021_inclination}.

 The impact of the star-dust geometry is, to first order, to modulate the slope of the attenuation curve such that galaxies with clumpier geometries, in which stars are increasingly uncoupled from the dust, have shallower (greyer) attenuation curves. With simple analytic geometries, \cite{witt_gordon_1996, witt_gordon_2000} highlighted the impact of the star-dust geometry on the resulting dust attenuation curve, where an increase in the inhomogeneity of the ISM structure results in increasingly shallow attenuation curves. Similarly, \cite{tuffs_2004_clumpiness} presented an analytical model that accounted for attenuation in discrete regions (bulge, thin disk, thick disk) targeting young and old stellar populations separately, with an additional 'clumpy' attenuation component that impacts young stars still in their birth clouds. Numerical experiments allowing for more complex and self-consistent star-dust geometries from idealized \citep[e.g.][]{hayward_smith_2015, natale_2015} and cosmological hydrodynamical simulations \citep[e.g.][]{trayford_2017, narayanan_2018_atten} amplified these findings. As shown in \cite{narayanan_2018_atten} with hydrodynamical cosmological zoom-in simulations, the steepest normalized dust attenuation curves are found in galaxies with a significant fraction of young stars (dominating the emitted UV flux) obscured by dust but old stars (dominating the optical flux) that are not obscured. A more mixed geometry (i.e., more old stars obscured by dust or more young stars decoupled from dust) will flatten the attenuation curve from this extreme limit. Observational studies \citep[e.g.][]{tress_2018} have also attributed variations in UV bump strength, and its inferred correlation with the total-to-selective extinction ratio (R$_V$), to the variations in star-dust geometry from galaxy to galaxy. 
 
 Observational constraints on the degree to which stars and dust are spatially coupled are difficult to obtain but recent work by \cite{leslie_2018_clumpiness} and \cite{vandergiessen_2022_dust_clumpiness} using the \cite{tuffs_2004_clumpiness} model show that galaxies in both SDSS and COSMOS exhibit significant 'clumpiness' fractions, such that young stars are heavily attenuated and the overall attenuation curves are shaped by this non-uniformity in dust geometry. Pa$\beta$ and H$\alpha$ emission line ratios, which can be used to infer dust attenuation, have been shown to lie outside the expected relation when measured in resolved galaxies with prominent dust lanes, as demonstrated in \cite{cleri_2020_pabeta}, indicating the need to accommodate the star-dust geometry in dust attenuation measurements.  
 
 \subsection{How Do We Account for Geometry in Dust Attenuation Models?}
The most common way to model the impact of the star-dust geometry on a galaxy's dust attenuation curve is to allow a variable UV-optical slope: 

\begin{equation} \label{eq:slope}
\begin{aligned}
    \tau_{\lambda} = -{{\rm ln}(I / I_0)} \\
    S_\mathrm{UV-Opt} \equiv A_{1500} / A_{V}
\end{aligned}
\end{equation}

where $\tau_{\lambda}$ is the opacity calculated via the ratio of $\mathrm{I}$, the composite spectrum, to ${I_0}$, intrinsic stellar spectrum, and S$_\mathrm{UV-Opt}$ is the ratio of attenuation at $1500 \AA$ to the attenuation in the V-band. $\tau_{\lambda}$ relates to the attenuation as $A_{\lambda} \approx 1.086 \: \tau_{\lambda}$.

For example, in studies like \citet{noll_2009} and \citet{salim_2018} the authors allow the UV-optical attenuation curve slope to vary by multiplying the \cite{calzetti_2000} curve with a power-law term centered at the V band:
 
 \begin{equation}\label{eq:power-law}
     k(\lambda)_{\rm variable} \propto k(\lambda)_{\rm Calzetti}  \left(\frac{\lambda}{5500\AA}\right)^{\delta}
 \end{equation}
 
where $k(\lambda)_{variable}$ is the model attenuation curve, $k(\lambda)_\mathrm{Calzetti}$ is the base \cite{calzetti_2000} curve, and $\delta$ is the variable power-law slope (hereafter referred to as the UV-optical slope). Negative (positive) values of the slope term give attenuation curves that are steeper (shallower) than the Calzetti curve at $\lambda < 5500 \AA$, allowing the attenuation curve to model galaxy-to-galaxy variations in attenuation curves manifested by, e.g., varying star-dust geometries. We show an example of this model in the top panels of Figure \ref{fig:frac_vs_slope_shape}, where we plot SEDs and attenuation curves generated from simple stellar populations assuming a \cite{calzetti_2000} curve with a variable power-law slope. A modest range in the $\delta$ power-law slope parameter space results in a wide range of SEDs and attenuation curves even for fixed values of $\tau_V$. 

\begin{figure}
    \centering
    \includegraphics[width=0.49\textwidth]{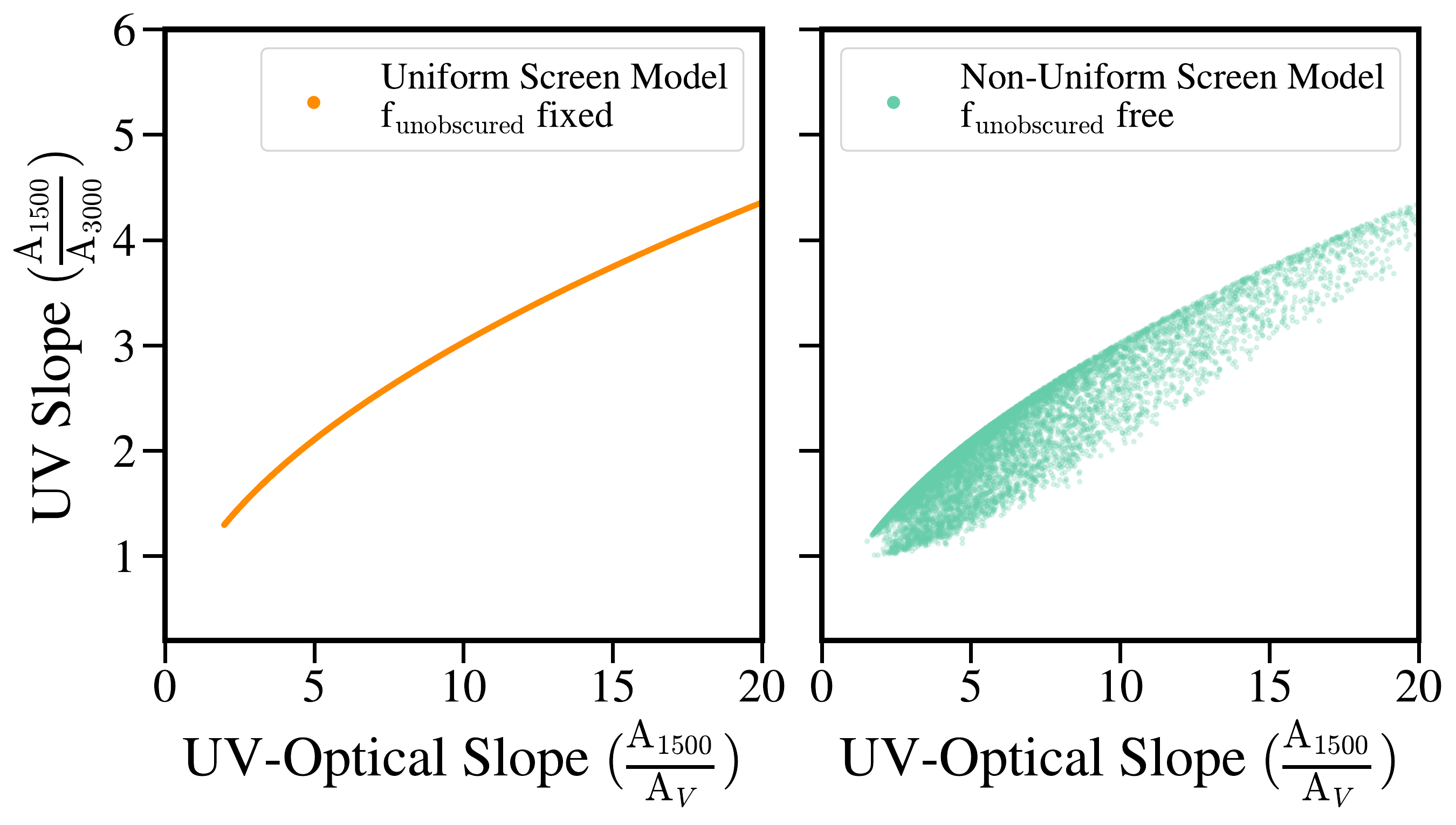}
    \caption{Comparison of the UV slope and UV-optical slope parameter space probed by the uniform and non-uniform screen attenuation curve models. Plotted are 10000 draws from the model parameter prior distributions.}
    \label{fig:UV_slope_space}
\end{figure}

We consider this implementation of the Calzetti curve model a 'uniform screen,' since all stellar light is assumed to be attenuated equally by the same optical depth regardless of stellar age. While the variable UV-optical slope model enables us to approximate the complex star-dust geometries, it is limited in cases where the attenuation curve deviates from, e.g., a power-law function. We explore in the following section an alternative model that can accommodate more diverse attenuation curve shapes by removing the limitation of a power-law model.
 
 \subsection{Is A Power-law Slope Enough?} 
 
 As discussed above, the effect of star-dust geometry can be approximated by a variable attenuation curve slope, commonly parameterized as a power-law deviation from the fiducial \citet{calzetti_2000} curve. What are the limitations of such a model? To explore this question, we introduce a parameter, called f$_\mathrm{unobscured}$, that allows deviations from a uniform screen model. This parameter controls the fraction of the composite stellar SED that is attenuated by the dust attenuation curve; we can think of ($1 - \mathrm{f}_\mathrm{unobscured}$) as a covering fraction in which nonzero values of f$_\mathrm{unobscured}$ result in a non-uniform screen model. The model composite spectrum with this parameter becomes:
 
  \begin{equation}\label{eq:spectrum}
     I = I_0 k_{\lambda} (1 - f_\mathrm{unobscured}) + I_0 f_\mathrm{unobscured}
 \end{equation}
 
 To see the impact of this parameter on the attenuation curve model, we revisit Figure \ref{fig:frac_vs_slope_shape}. In the bottom panel, we plot the SED and attenuation curves for a non-uniform screen model where we vary the fraction of unobscured stellar light. For simplicity, we plot the attenuation curves for two fixed values of $\delta$, but in practice the slope would be allowed to vary as in the top panels. For a fixed $\delta$ slope, increasingly larger fractions of unattenuated light result in greyer attenuation curves that flatten out at small wavelengths.
 
 For the uniform screen model parameterization adopted in this paper \citep{kc_law}, the shape of the attenuation curve is tied to the shape of the \citet{calzetti_2000} curve plus a 2175$\AA$ bump feature. Tying the attenuation curve shape to the Calzetti curve represents a restriction on our model; the slope in one wavelength range of the attenuation curve is tied to the slope at $5500 \AA$. This is shown explicitly in Figure \ref{fig:UV_slope_space}, where we plot the prior distributions for the UV slope and the UV-optical slope for the uniform screen and non-uniform screen models. The uniform screen model restricts the possible shapes of the attenuation curves. In contrast, the non-uniform screen allows for a range of UV slope values for fixed values of UV-optical slope. Variations in f$_\mathrm{unobscured}$ result in attenuation curves that are distinctly different than the curves generated by a uniform screen hinting that a more complex distribution of stars and dust changes the attenuation curve in ways that cannot easily be captured by varying a power-law slope alone.

The addition of the f$_\mathrm{unobscured}$ parameter to a variable power-law slope model enables physically motivated flexibility without needing to make choices about the exact configuration of the star-dust geometry. In our analysis, we opt to remain agnostic about the specific configuration of the star-dust geometry by instead modeling the aggregate effect of the geometry on the attenuation curve. The goals of this paper are to show that the addition of f$_{\rm unobscured}$ introduces a necessary degree-of-freedom into the attenuation curve model such that we are better able to reproduce the attenuation curve of galaxies from a cosmological simulation.

\begin{figure}
    \centering
    \includegraphics[width=0.48\textwidth]{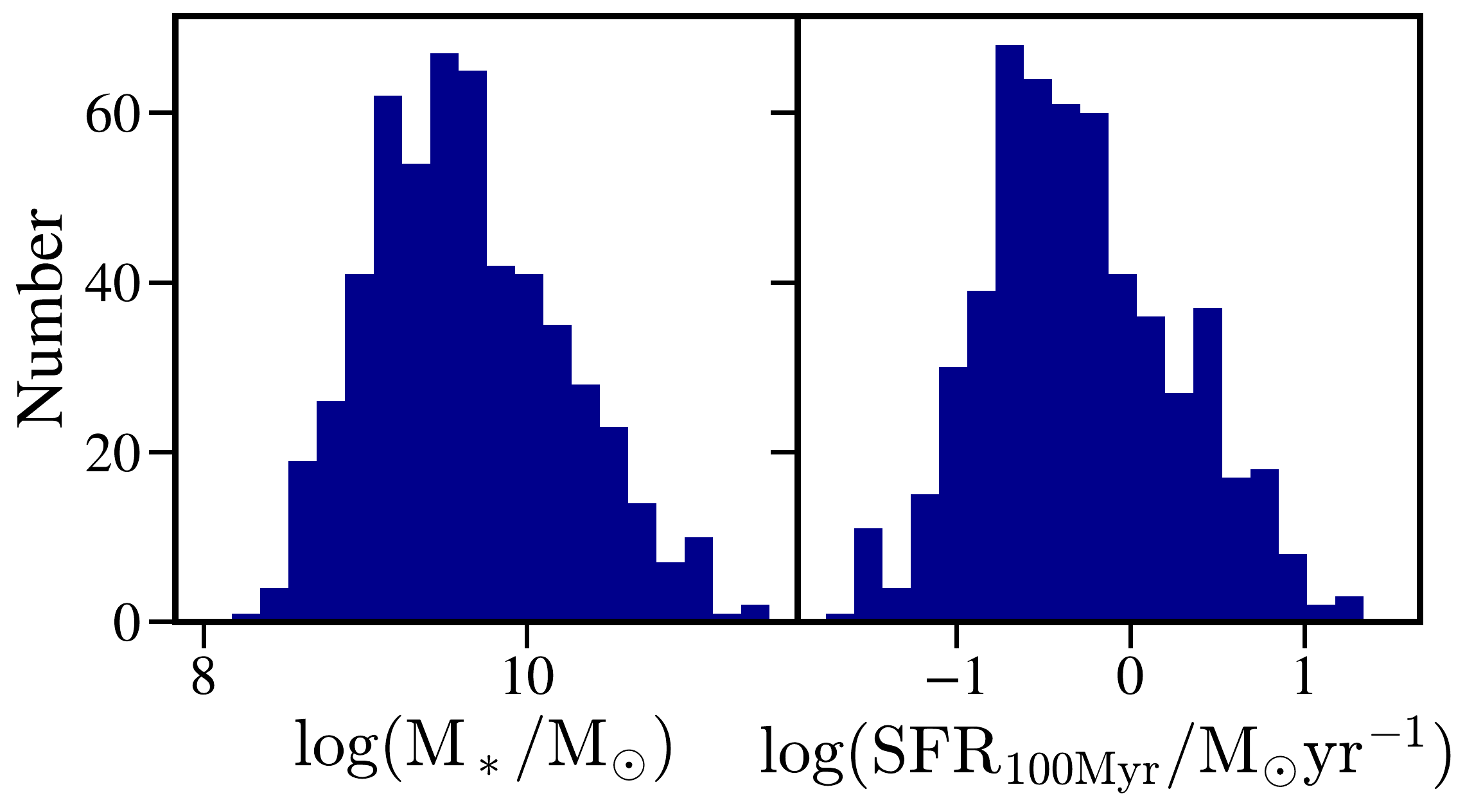}
    \includegraphics[width=0.48\textwidth]{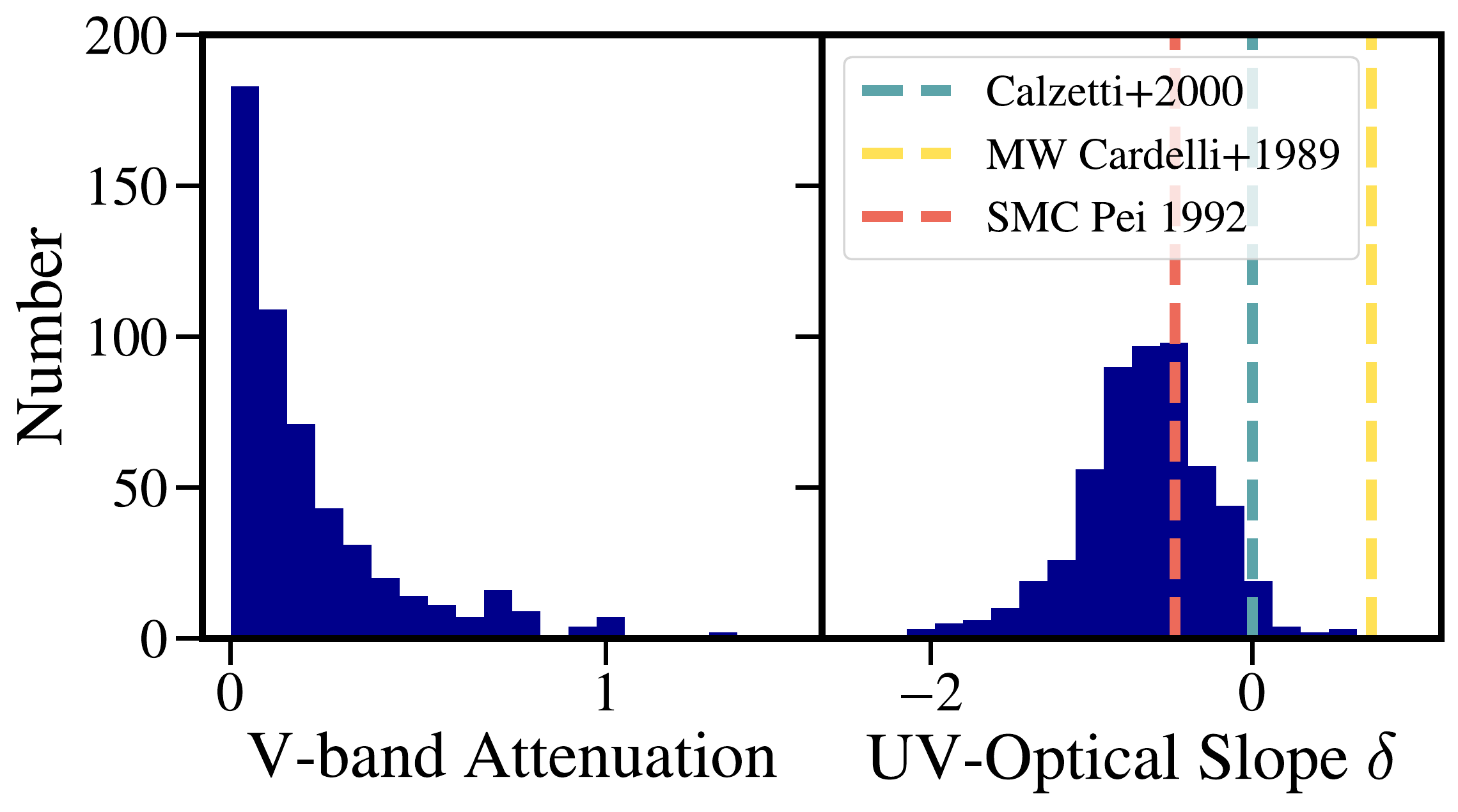}
    \includegraphics[width=0.48\textwidth]{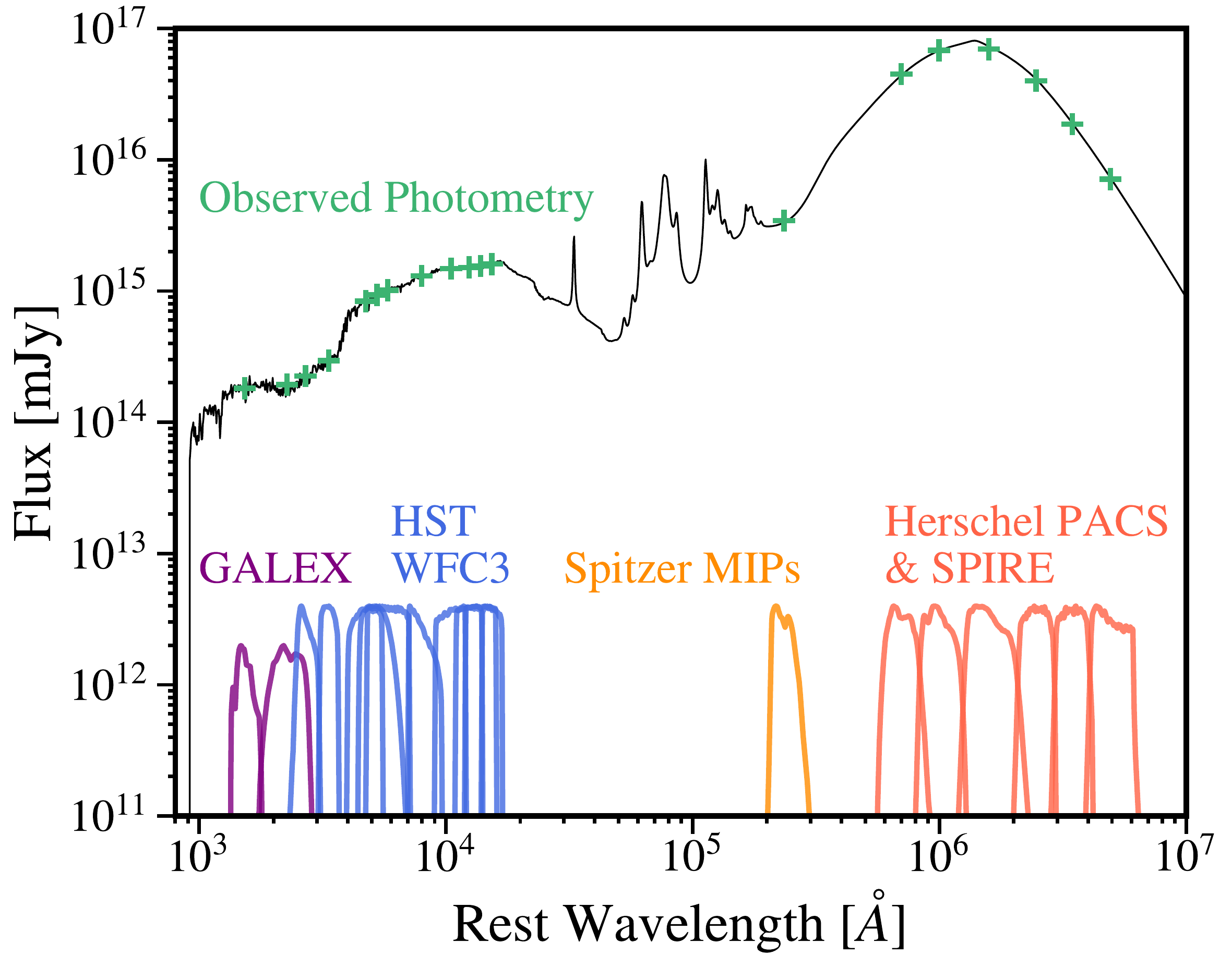}
    \caption{\textbf{Top}: Distribution of the simulated galaxy physical properties stellar mass and SFR. \textbf{Middle} Distribution of dust attenuation properties for the simulated galaxies. \textbf{Bottom}: Example {\sc powderday} SED with the $19$ broadband filters used in this analysis overlaid. We neglect coverage of the PAH features in the mid-IR, as well as fitting for the PAHs in the SED fits, as these are template based and not self-consistently modeled. }
    \label{fig:sed}
\end{figure}

\section{Numerical Methods} \label{sec:methods}

\subsection{Overview}

In order to understand the current efficacy of dust attenuation modeling, we employ galaxies from the {\sc simba} hydrodynamical cosmological simulation coupled with post-processing radiative transfer to produce synthetic SEDs that we model with {\sc prospector}, a flexible SED modeling code. Below we describe the {\sc simba} galaxy formation model, the 3D dust radiative transfer, and the {\sc prospector} SED models used to fit the {\sc powderday} synthetic SEDs. 

\subsection{{\sc simba} Cosmological Simulation}\label{sec:simba}

We utilize the (25/h Mpc)$^3$ {\sc simba}\footnote{\url{simba.roe.ac.uk}} volume with $2 \times 512^3$ particles, resulting in a baryonic mass resolution of  $1.4 \times 10^6$ M$_{\odot}$. The {\sc simba} galaxy formation model is a hydrodynamical cosmological simulation based on the {\sc gizmo} gravity and hydrodynamics code \citep{hopkins_2015_gizmo} and assumes \textit{Planck} cosmology. The fundamental models describing galaxy formation and evolution include (1) an H$_2$-based star formation rate
(SFR); (2) a chemical enrichment model that tracks 11 elements from Type II supernovae (SNe), Type Ia SNe, and asymptotic giant branch (AGB) stars; (3) subresolution models for stellar feedback including contributions from Type II SNe, radiation pressure, and stellar winds; (4) subresolution models for feedback via active galactic nuclei (AGN) including a two-phase jet (high accretion rate ) and radiative (low accretion rate) model; and (5) a self-consistent dust model \citep{li_2019_dust}, in which dust is produced by condensation of metals ejected from SNe and AGB stars and is allowed to grow by metal accretion, and be destroyed by thermal sputtering and astration processes. 

\begin{figure*}[t]
    \centering
    \includegraphics[width=0.48\textwidth]{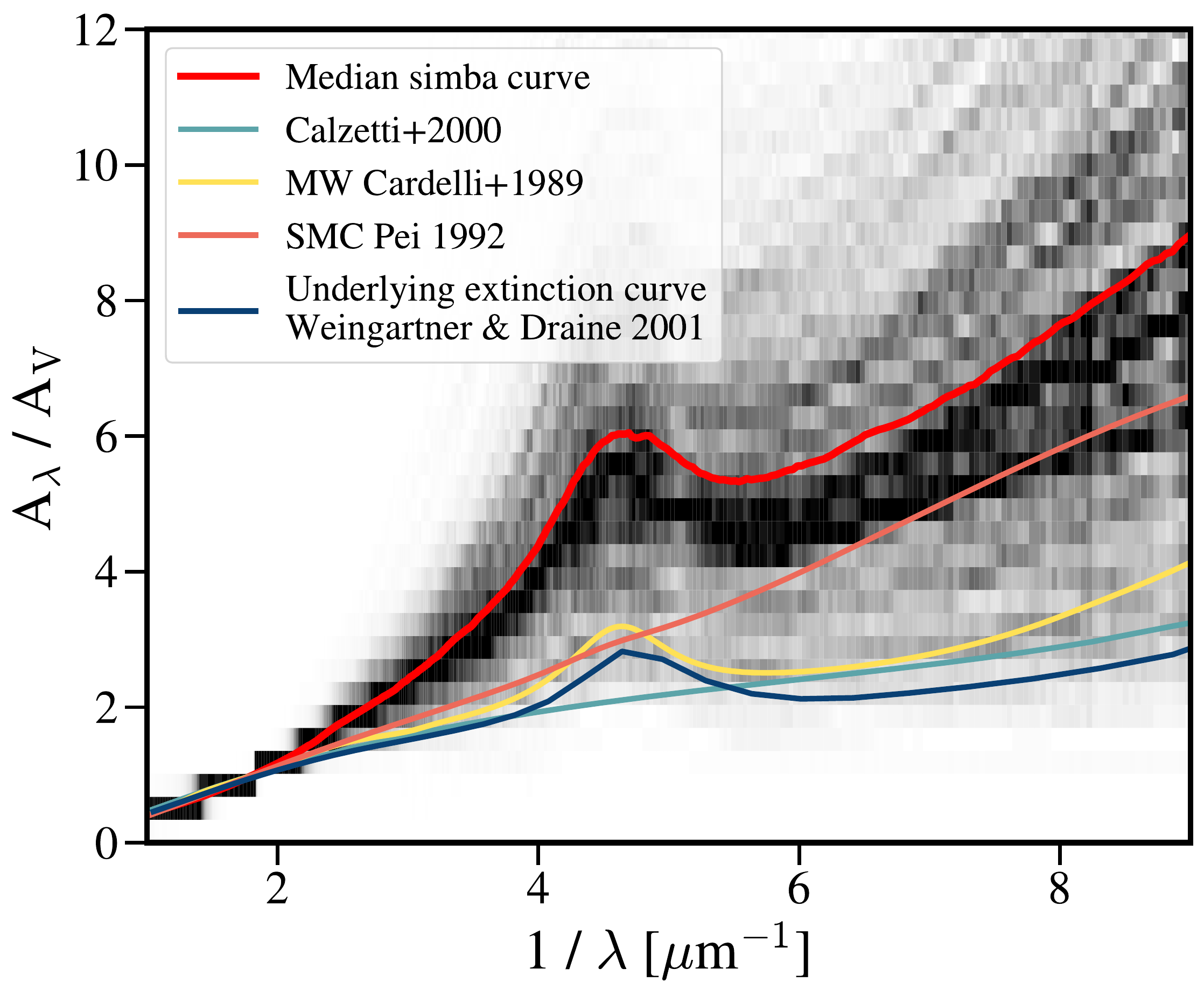}
    \includegraphics[width=0.48\textwidth]{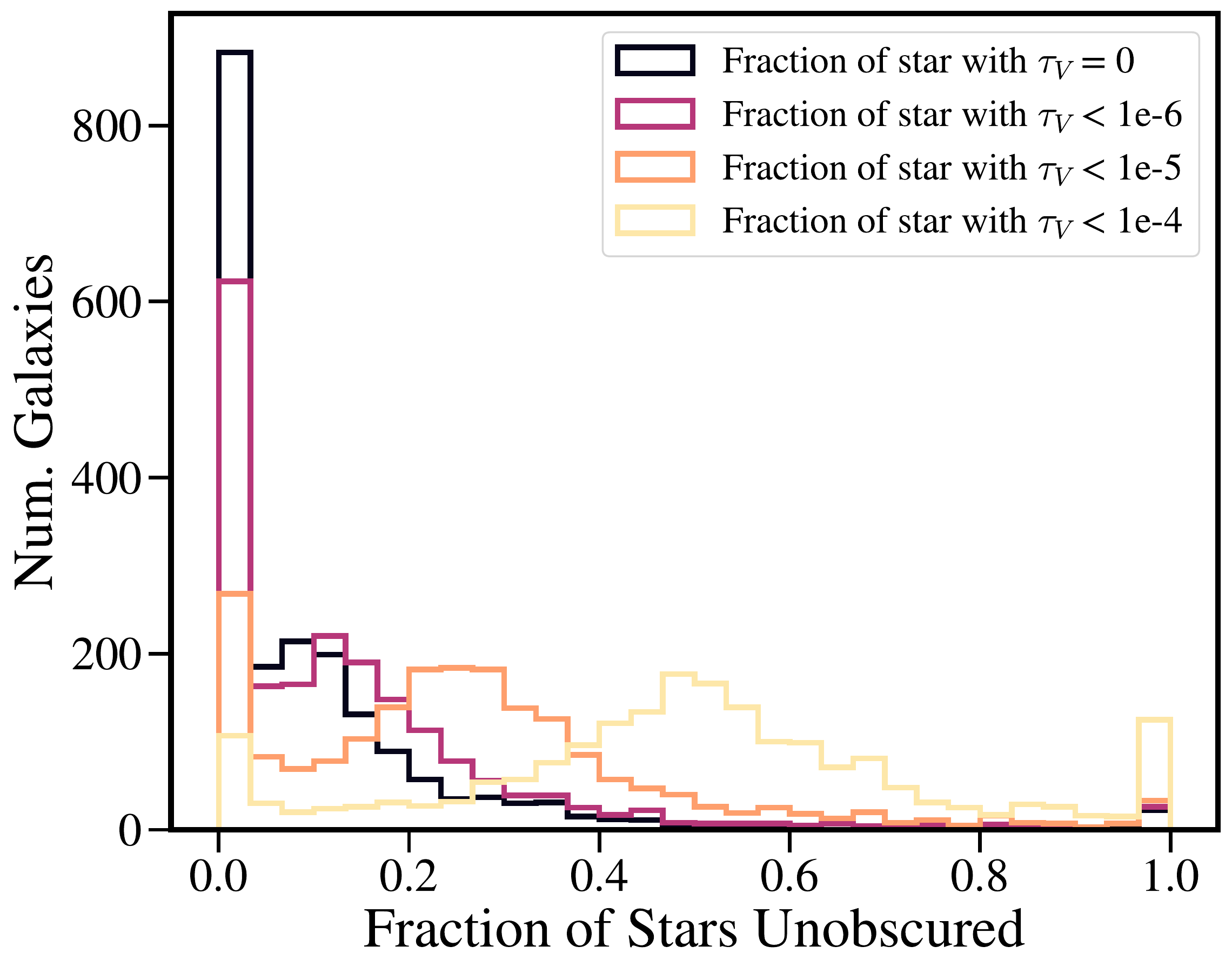}
    \caption{ \textbf{Left}: Distribution of the $z = 0$ simulated attenuation curves for star-forming galaxies, where the heatmap shows the density of curve parameter space, with the median curve shown in red, along with the underlying extinction curve \citep{weingartner_draine} and three curves from the literature: the average SMC extinction curve \citep{pei_1992_smc}, the average MW extinction curve \citep{ccm_1989_mw_ext}, and the \cite{calzetti_2000} curve. \textbf{Right}: Fraction of stars unobscured by dust in the simulated {\sc simba} galaxies as a function of the definition of 'no dust,' i.e. the limit on V band optical depth. This was computed by calculating the line-of-sight column dust density for each star in $\sim$1000 galaxies for one sight-line.}
    \label{fig:simba_props}
\end{figure*}

We identify galaxies in the $z = 0$ {\sc simba} snapshot using the {\sc caesar} \citep{caesar} 6-D friends-of-friends galaxy finder based on the number of bound stellar particles in a system: a minimum of 32 stellar particles defines a galaxy. We then select galaxies with at least one star particle formed in the last 100 Myr to provide a star formation rate for comparison purposes. Our galaxy sample is shown in the top and middle panels of Figure \ref{fig:sed}, where we show the distribution of stellar mass, SFR, and attenuation curve properties.

\subsection{{\sc Powderday} 3D Dust Radiative Transfer}\label{sec:pd}

\subsubsection{Generating Synthetic SEDs}
We use the radiative transfer code {\sc powderday} \citep{powderday} to construct synthetic SEDs for the {\sc simba} galaxies selected in \ref{sec:simba}. {\sc powderday} wraps the stellar population synthesis code {\sc fsps} \citep{fsps_1, fsps_2}, the 3D dust radiative transfer code {\sc hyperion} \citep{robitaille_2011_hyperion} and {\sc yt} \citep{yt}. We use the stellar ages and metallicities as returned from the cosmological simulation to generate the dust free SEDs for the star particles within each cell, assuming a \cite{kroupa_2002_imf} stellar IMF and the {\sc mist} stellar isochrones \citep{choi_2016_mist, dotter_2016_mist}. We neglect contributions from active galactic nuclei (AGN) and nebular emission, though AGN emission can be enabled in {\sc powderday} \citep{sharma_2021}. The stellar SEDs are then propagated through the dusty ISM, derived from the on-the-fly self-consistent model of \cite{li_2019_dust}. Due to the limited mass resolution of the simulation, birth clouds are not modeled self-consistently; as such, in our later analysis of SED fitting, we test only the diffuse dust component of our galaxies. This diffuse dust is modeled as a carbonaceous and silicate mix following \cite{draine_infrared_2007}, with the \cite{weingartner_draine} size distribution and the \cite{draine_03} re-normalization relative to hydrogen. Though polycyclic aromatic hydrocarbons (PAHs) are not modeled explicitly in {\sc simba}, their emission is included following the \cite{robitaille_pahs} model in which PAHs are assumed to occupy a constant fraction of the dust mass (here, modeled as grains with size a $< 20$ $\AA$) and occupying 5.86\% of the dust mass. The dust emissivities follow the \cite{draine_infrared_2007} model, though are parameterized in terms of the mean intensity absorbed by grains, rather than the average interstellar radiation field as in the original \citeauthor{draine_infrared_2007} model. 

The radiative transfer propagates through the dusty ISM in a Monte Carlo fashion using {\sc hyperion}, which follows the \cite{lucy_rt} algorithm in order to determine the equilibrium dust temperature in each cell. We iterate until the energy absorbed by 99\% of the cells has changed by less than 1\%. Note that while we assume the \cite{weingartner_draine} extinction curve in every cell for each galaxy, the effective attenuation curve is a function of the star-dust geometry, and therefore varies from galaxy to galaxy \citep{attenuation_araa}. We then sample broadband photometry in $19$ bands, including \textit{GALEX}, HST, \textit{Spitzer}, and \textit{Herschel}, from the synthetic {\sc powderday} SEDs, assuming a uniform signal-to-noise ratio of $30$ in all bands. An example {\sc powderday} SED is shown in the bottom panel of Figure \ref{fig:sed}. 

\subsubsection{Calculating Attenuation Curves}

We calculate the attenuation curves for each galaxy following Equation \ref{eq:slope}. In the left panel of Figure \ref{fig:simba_props}, we show the distribution of simulated galaxy attenuation curves. The heatmap shows the density of curve parameter space and the red line denotes the median attenuation curve. Though the \cite{weingartner_draine} extinction curve, shown in dark blue, is the same for all galaxies, the effective attenuation curves cover a wide range of parameter space. This wide range is primarily driven by differences in the star-dust geometry and the stellar age distributions.

In the context of the discussion in \S \ref{sec:toy_models}, one way to parameterize the star-dust geometry is shown in the right panel of Figure \ref{fig:simba_props}, where we plot the distribution of f$_{\rm unobscured}$, defined here as the fraction of stars in each galaxy that are unobscured by dust, for different $\tau_{\lambda}$ limits defining 'unobscured.' We calculate f$_{\rm unobscured}$ by computing the line-of-sight dust column density for each star particle in the simulated galaxies over one sight-line. Though this is not exactly how f$_{\rm unobscured}$ is defined in \S \ref{sec:toy_models}, this paints an idea of how the stars and dust particles are coupled in our simulated galaxies. 

\subsection{{\sc Prospector} SED Modeling}\label{sec:prosp}

In order to model the dust attenuation of the {\sc simba} galaxies, we use the flexible state-of-the-art SED modeling code {\sc prospector} \citep{leja_deriving_2017, prosp}. Similar to {\sc powderday}, {\sc prospector} is based on the stellar population synthesis code {\sc fsps} to generate simple stellar populations and convolved with the assumed star formation history (SFH) model to generate a composite stellar spectra. We fit the {\sc powderday} SEDs sampled at 19 broadband filters shown in the bottom panel of Figure \ref{fig:sed}. {\sc Prospector} relies on {\sc dynesty} \citep{dynesty}, a dynamic nested sampler that estimates Bayesian posteriors and evidences, to sample model parameter space. In our analysis, we closely follow the models and priors outlined in \cite{lower_stellar_mass} including the stellar mass-stellar metallicity relation of \cite{gallazzi_ages_2005} and the 6 component nonparametric SFH using a flexible modified Dirichlet prior \citep{Betancourt_2013_dirichlet, leja_2019_nonpara} on the fractional stellar masses formed in each time bin. 

{\sc Prospector} includes several models for dust attenuation within the {\sc fsps} framework. Notably, parameters specifying the attenuated fraction of stellar light -- that is, the fraction of stars, both young ($< 10$ Myr old) and old, that are unobscured by dust -- can be varied. Meaning, {\sc prospector} enables the dust attenuation model to be more diverse than the traditionally assumed uniform screen model. Because we do not model birth clouds in either the {\sc simba} model (due to limited mass resolution) nor the {\sc powderday} radiative transfer, we exclude any explicit age-based differential attenuation such as the, e.g., \cite{charlot_fall_2000} model. We instead focus only on the diffuse dust component. The three model variations we consider in our analysis are described below:

\textbf{Fixed curve shape with uniform screen}: Here, we assume the \cite{calzetti_2000} model. The slope of this curve is fixed and we implement this curve as a uniform screen, meaning all stellar light is impacted by this attenuation. The \cite{calzetti_2000} curve does not have the 2175 $\AA$ bump feature. We allow the normalization (V band optical depth) to vary with a truncated Gaussian prior, centered at A$_V = 0$ with a width of $0.3$ and truncated at A$_V  = 2.5$.

\textbf{Variable curve shape with uniform screen}: Here, we use the attenuation curve parameterization of \cite{kc_law}, which is based on the above \cite{calzetti_2000} relation, modulated by a power-law slope and includes a variable 2175 $\AA$ feature where the strength of the bump is tied to the slope deviation from the \cite{calzetti_2000} curve, in the sense that steeper curves have larger bump strengths. This slope allowed is to vary, ranging from -1.8 to 0.3 with a uniform prior relative to the \cite{calzetti_2000} curve slope as in Equation \ref{eq:power-law}. The prior on V-band attenuation is the same as the fixed curve slope model. 

 \begin{figure*}[t]
    \centering
    \includegraphics[width=\textwidth]{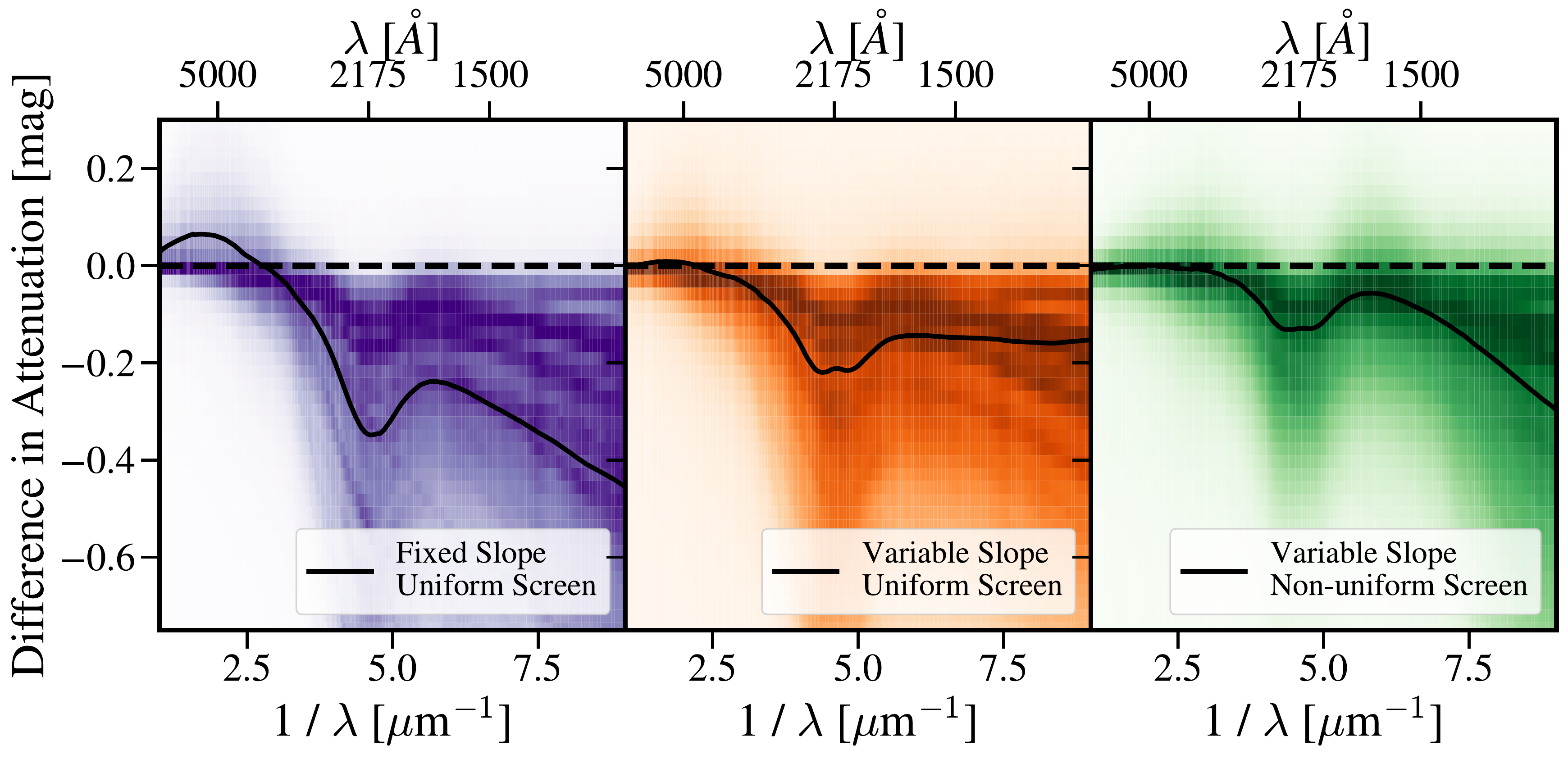}\\
    \includegraphics[width=0.48\textwidth]{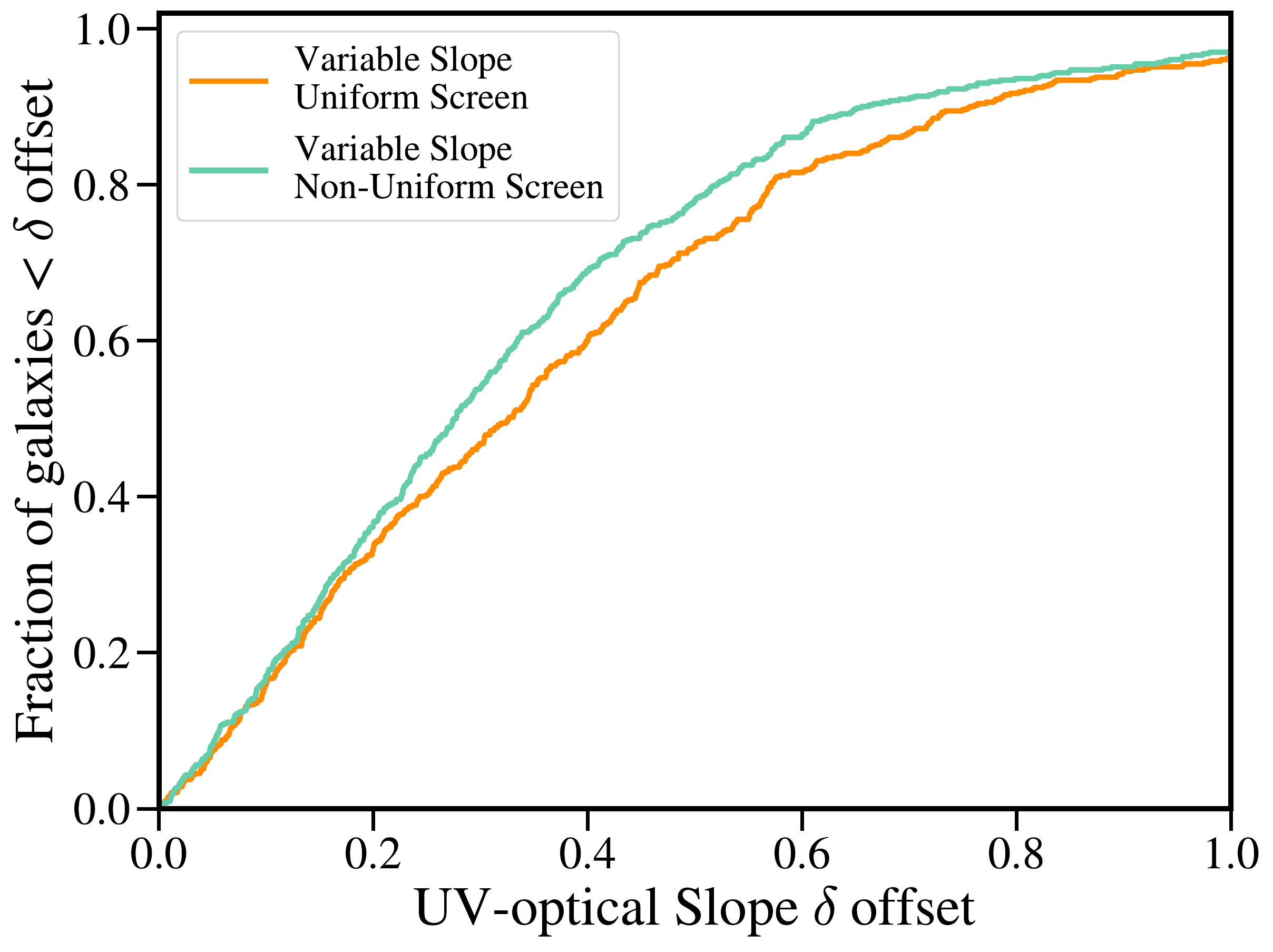}
    \includegraphics[width=0.48\textwidth]{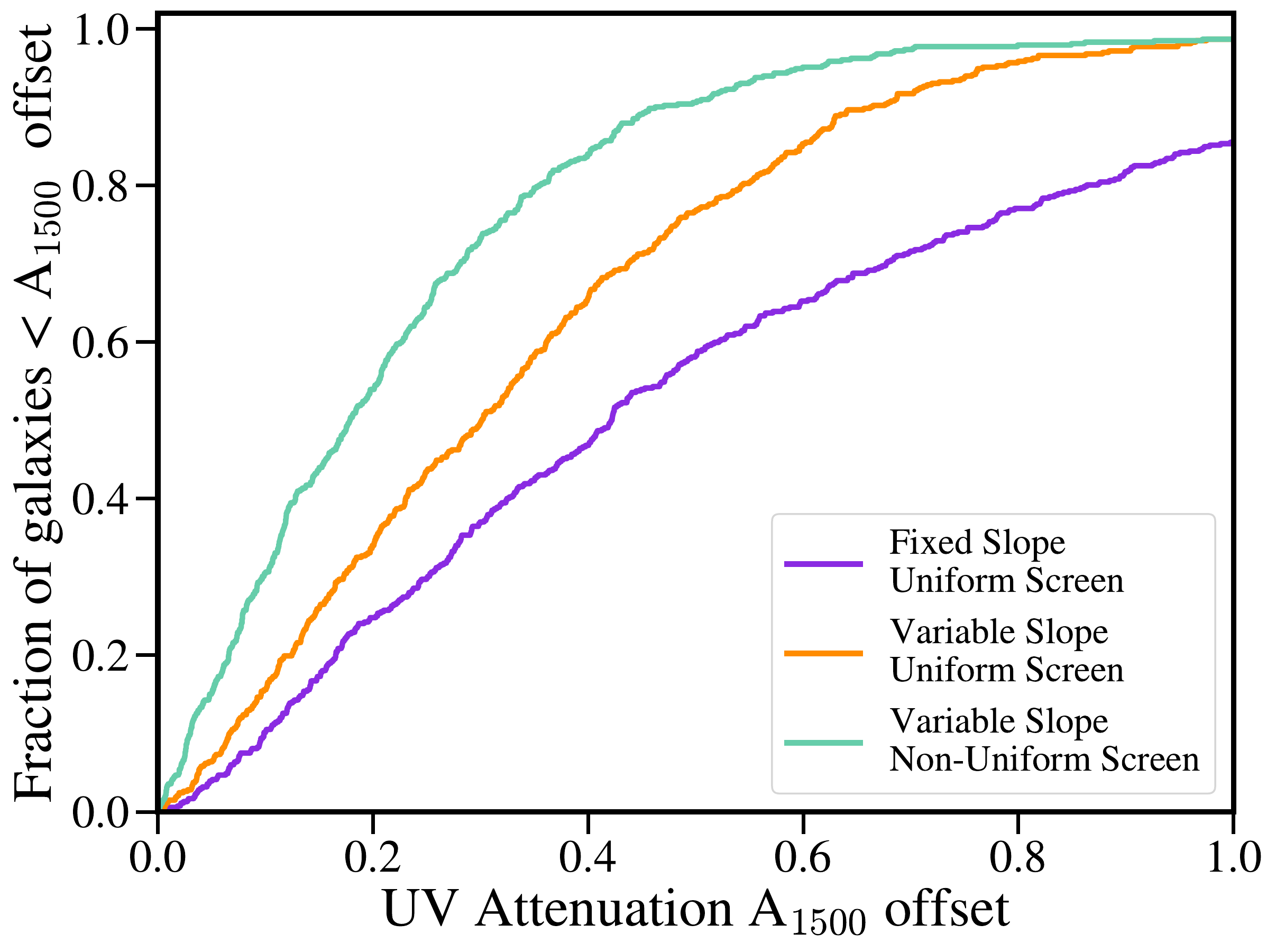}
    \caption{\textbf{Top}: Heatmaps showing the distribution of attenuation curve offsets (model curve - true curve) for each attenuation curve model. Darker regions of the heatmap indicate higher density in that bin. The solid lines are the median offsets for that model for the sample of $\sim$550 galaxies. Negative values result from under-estimates in the attenuation at those wavelengths. \textbf{Bottom}: Cumulative distribution functions showing the magnitude of offsets for the UV-optical slope (left) and attenuation at $1500 \AA$ (right).}
    \label{fig:attenuation_curves}
\end{figure*}

 \textbf{Variable curve shape with non-uniform screen}: This model is similar to the above variable slope model except we also allow the covering fraction of dust to vary. This parameter, f$_{\rm unobscured}$, modulates the covering fraction of the diffuse dust and allows us to model non-uniformity in the star-dust geometry. Though the power-law slope and the fraction of unobscured stellar light are expected to be somewhat degenerate, the use of a Bayesian sampler means we can fully map these covariances and our uncertainties on the model parameters will be an accurate reflection of model degeneracies. The priors for the power-law slope and V-band normalization are the same as above while the prior on f$_{\rm unobscured}$ is uniform from $0$ to $1$.

\section{Results}\label{sec:results}

\subsection{Recovering Dust Attenuation Curves}\label{sec:recovering_attn}

As a first exercise, we fit the SEDs in the $z=0$ {\sc simba} volume using the three aforementioned dust attenuation models. In the top panel of Figure \ref{fig:attenuation_curves}, we show the distribution of offsets between the inferred attenuation curves and the true attenuation curves across the UV and optical wavelength range probed by the broadband photometry. We plot these offsets as a function of inverse wavelength to focus on the UV portion of the curve, as the three models converge around 5500 $\AA$ to a median offset of $\sim$0. We generate the inferred attenuation curve from the marginalized parameter values. As we expect degeneracies between model parameters, namely metallicity, the attenuation power-law slope, and f$_\mathrm{unobscured}$, we sample the full parameter posteriors to generate the attenuation curves instead of using the maximum likelihood values. The median offset from the true attenuation curve for 530 galaxies, represented by the black line in Figure \ref{fig:attenuation_curves}, is smallest for the non-uniform screen model at all wavelengths greater than $1500 \AA$, where we have available UV data.

\begin{figure}
    \centering
    \includegraphics[width=0.5\textwidth]{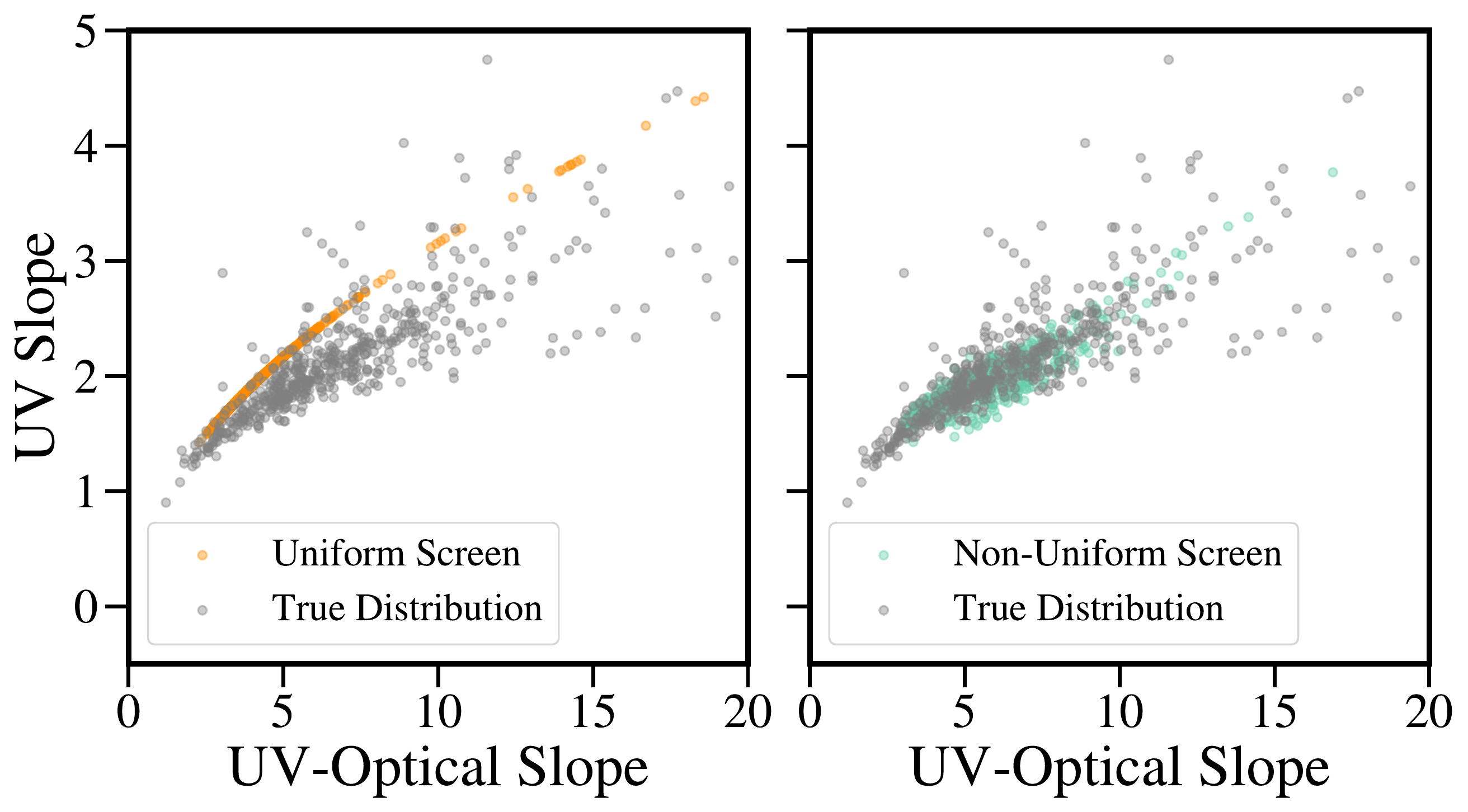}
    \caption{Same as Figure \ref{fig:UV_slope_space} but this time the slopes inferred from each model are plotted for the simulated galaxies. The distribution of slopes from the simulated attenuation curves are shown in gray.} 
    \label{fig:UV_slope_space_results}
\end{figure}

We explore this further by examining which aspects of the curve see improvements in modeling in the bottom two panels of Figure \ref{fig:attenuation_curves}, where we plot the magnitude of offsets for the UV-optical slope and attenuation at $1500 \AA$. As above, we calculate the offsets between the marginalized model parameters and the true values. The slope here is defined as the power-law deviation from the UV-optical slope of the \cite{calzetti_2000} curve. While the decrease in bias of the curve slope between the uniform screen and non-uniform screen models is only marginal, there is a significant decrease in offset between the inferred and true attenuation in the UV for the non-uniform screen model. This implies that while the variable power-law slope uniform screen model can infer the shape of the curve between $1500 \AA$ and the optical regime, it significantly under-estimates the attenuation in the UV with a median bias of $-0.3$ magnitudes compared to $-0.17$ for the non-uniform screen model. We interpret this as a consequence of the limited capability of a power-law transformation of the \cite{calzetti_2000} curve to match the diversity of the {\sc simba} attenuation curves. Supporting this, in Figure \ref{fig:UV_slope_space_results}, we plot the the UV slope inferred by the two variable slope models as a function of UV-optical slope (similar to Figure \ref{fig:UV_slope_space}) which demonstrates the inability of the uniform screen model to cover the parameter space occupied by the {\sc simba} galaxies (shown in gray). The tight relation between the two slopes for the uniform screen model is a consequence of the fact that the shape of the attenuation curve is allowed to vary only as a power-law deviation from the \citealt{calzetti_2000} curve shape.

Lastly, we note that all three attenuation curve models also appear to significantly under-estimate the strength of the 2175 $\AA$ bump feature. As shown in the left panel of Figure \ref{fig:simba_props}, the average simulated attenuation curve features the 2175 $\AA$ bump, as does the underlying extinction curve used to generate the {\sc powderday} SEDs. Because the attenuation curve model with a fixed slope is based on the \cite{calzetti_2000} curve, this deficit in the $2175 \AA$ bump strength is to be expected as the \citealt{calzetti_2000} curve does not include this feature. The two variable slope models are based on the \cite{kc_law} curve, where the strength of the bump feature is a function of the slope of the curve. \cite{kc_law} found that galaxies exhibiting steeper attenuation curves had smaller bump features. The bump strengths the authors measured are factors of 2-3 smaller than the Milky Way bump strength for sight-lines measured by \cite{valencic_2004_MW_bump_strength}. Thus, compared to the {\sc powderday} attenuation curves, the bump strengths of the \cite{kc_law} model at similar slopes are 1.5$\times$ smaller. Therefore, these deficits in the inferred $2175 \AA$ bump feature are also expected.

\begin{figure}[t]
    \centering
    \includegraphics[width=0.5\textwidth]{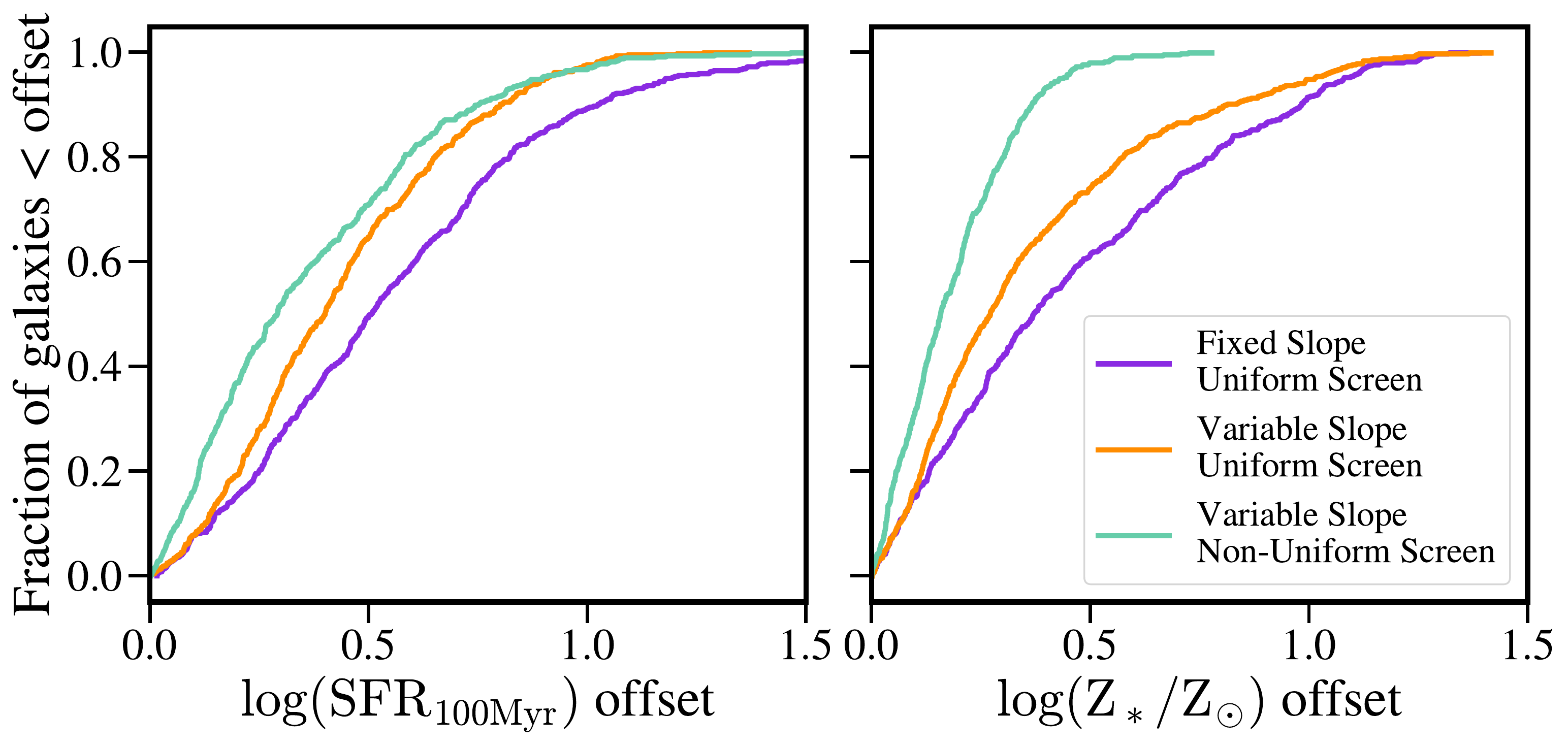}
    \caption{Cumulative distribution functions for the magnitude of offsets between the star formation rates (SFRs, left) and stellar metallicity (right) inferred from the three attenuation curve models and the true values. The SFRs are  averaged over the last 100 Myr.}
    \label{fig:sfr}
\end{figure}

\begin{figure}[t]
    \centering
    \includegraphics[width=0.5\textwidth]{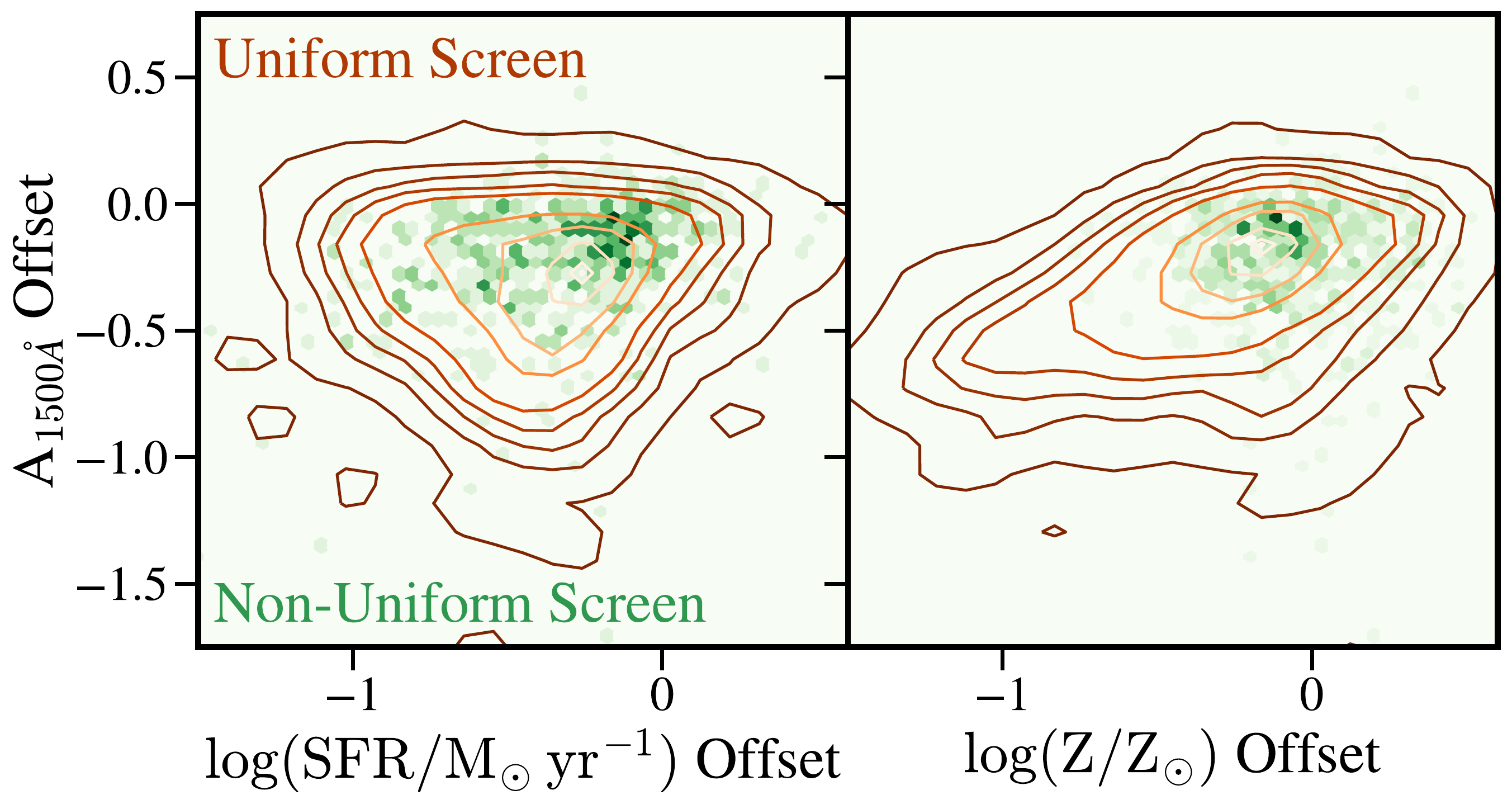}\\
    \includegraphics[width=0.5\textwidth]{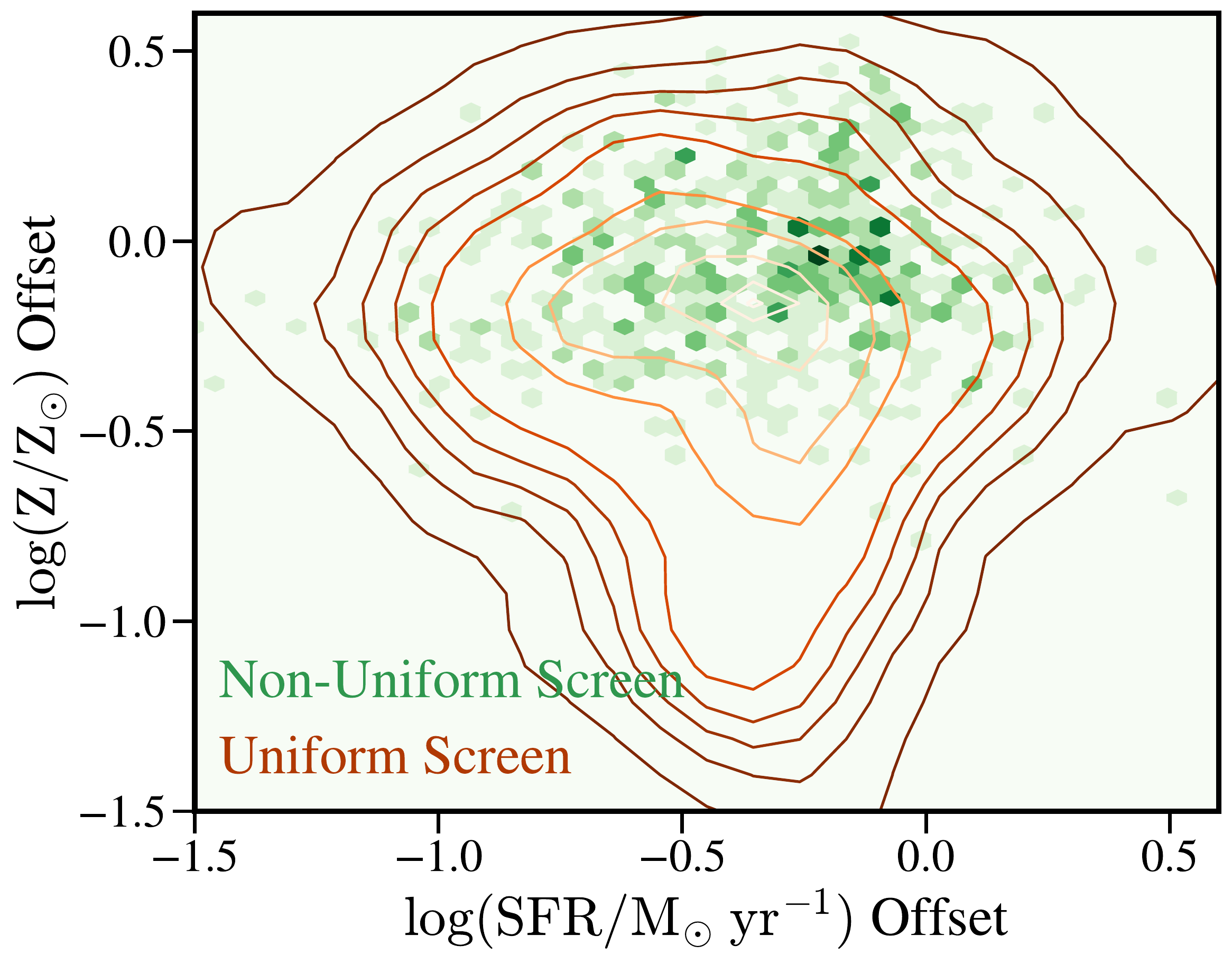}
    \caption{\textbf{Top}: Plot of the UV attenuation offsets as a function of SFR offset and stellar metallicity offset for the two variable slope attenuation models. The green hexbins show the non-uniform screen model while the orange contours show the uniform screen model. \textbf{Bottom} Plot of stellar metallicity offsets as a function of SFR offset.}
    \label{fig:a1500_sfr_Z}
\end{figure}

\subsection{Inferred Galaxy Properties}\label{sec:recovering_props}

A challenging aspect of SED modeling is the necessity to model each component (stars, metallicity, dust, etc.) simultaneously and untangle the numerous degeneracies associated with each aspect to infer the physical properties of a galaxy. From the previous section, we demonstrated that the non-uniform screen attenuation model successfully inferred the shape of the true dust attenuation curve for a majority of galaxies (with a median bias in the near-UV regime of 0.16 magnitudes). 

How does this translate to the model's ability to infer the physical properties of these galaxies? In Figure \ref{fig:sfr} we plot the cumulative distribution of offsets for two inferred properties: the galaxy SFR and stellar metallicity for the three attenuation models. For both the SFR and galaxy metallicity, the increased accuracy of the modeled attenuation curves does translate to a modest increase in accuracy.  This is especially true for the stellar metallicity that sees a decrease in bias of $0.18$ dex on average. The SFRs are modestly improved with a decrease in bias of $0.12$ dex. To relate these improvements in physical properties to the attenuation curve model improvements, in Figure \ref{fig:a1500_sfr_Z} we plot the offset in UV attenuation as a function of SFR offset and stellar metallicity offset and for SFR and metallicity together. For galaxies fit with the uniform screen model, a large bias in inferred SFR or metallicity is matched by a similarly large bias in the UV attenuation. The spread in offsets for all three properties imply that the uniform screen model struggles to model these SED components simultaneously - at most, only two properties can be accurately inferred.

As we explored in \cite{lower_stellar_mass}, the dominant source of uncertainty when inferring galaxy SFRs and stellar masses from SED modeling is the form of the assumed star formation history (SFH), namely that relatively simple models for the SFH can bias the inferred physical properties. In this paper, we vary only the dust attenuation model between SED fits and note only negligible differences in the inferred stellar masses between the three models, but modest improvements in the inferred SFR and significant improvements in the stellar metallicity, indicating that the ability to correctly infer the shape of the dust attenuation curve is vital in correctly measuring galaxy physical properties.

\section{Model Efficacy and Caveats}\label{sec:caveats}

Here, we address caveats to our analysis and the non-uniform screen model for dust attenuation. Namely, we highlight that while the attenuation curves inferred with the non-uniform model are more accurate on average, this model is still an approximation to the true star-dust geometry of a galaxy and cannot be used to draw conclusions about the details of the relative coupling of the stars and dust in that galaxy. Second, we use Bayesian evidence to determine the odds that either variable slope model is preferred over the other given the available data. Finally, we discuss that while the increased flexibility of the non-uniform screen model can be constrained with {\sc prospector}, far-infrared luminosity constraints are key to the results we have presented.

\subsection{Modeling the True Star-Dust Geometry}\label{sec:true_geom}

By adding a single parameter to a flexible slope dust attenuation curve model used in SED modeling, we have shown that we can reasonably reproduce the true attenuation curves of model galaxies, leading to improved estimates for galaxy physical properties such as the SFR. This parameter allows a non-zero fraction of stellar light to be unattenuated by dust in the diffuse ISM, resulting in attenuation curves that are functionally different from modulating the power-law slope alone. A non-uniform screen model accommodates attenuation curves that show deviations from a power-law slope due to some stars seeing different optical depths than other stars. Coupled with a variable power-law slope, we are better equipped to model galaxy-to-galaxy variations in star-dust geometries that manifest as changes to the shape of the dust attenuation curve.

However, this non-uniform screen model is still only an approximation of the impact of star-dust geometry on a galaxy's attenuation curve. For instance, a drawback of the our current implementation of f$_{\rm unobscured}$ is that we neglect any variation of the obscured fraction as a function of stellar age (and effectively wavelength).  The light-weighted fraction of unobscured stellar light will therefore be different than the mass-weighted fraction.  \cite{narayanan_2018_atten} demonstrated that allowing larger fractions of older stars to remain unobscured will result in steeper normalized dust attenuation curves while larger fractions of unobscured young stars will result in flatter attenuation curves. For the model presented in this study, we are unable to differentiate between the impact of unobscured old stellar populations vs. the impact of unobscured young stellar populations on the model attenuation curve. This means that we cannot determine   \textit{which} stars are unobscured and \textit{where} this decoupling happens within the galaxy. In essence, while f$_{\rm unobscured}$ is not explicitly the star-dust geometry, it aggregates the distribution of stellar obscuration fractions in a galaxy into an approximate bolometric covering fraction.

To overcome these model limits, future implementations of dust attenuation curves in SED models can benefit from a non-parametric treatment as is done for the SFH model in this work and developed in \citet{leja_deriving_2017} and \citet{leja_2019_nonpara}. A non-parametric dust attenuation model would consist of several piece-wise functions whose normalizations could vary independently, effectively accommodating any dust attenuation curve shape. Such a model, when properly constrained by full SED wavelength coverage and carefully chosen informative priors, would be less biased than the approaches taken in this paper since no assumptions regarding dust properties or star-dust geometries would be made. While the use of parametric models may make SED fitting results precise, as is also the case for SFH modeling, the limited nature of those models means the results will potentially be precise and wrong. 

\begin{figure}
    \centering
    \includegraphics[width=0.48\textwidth]{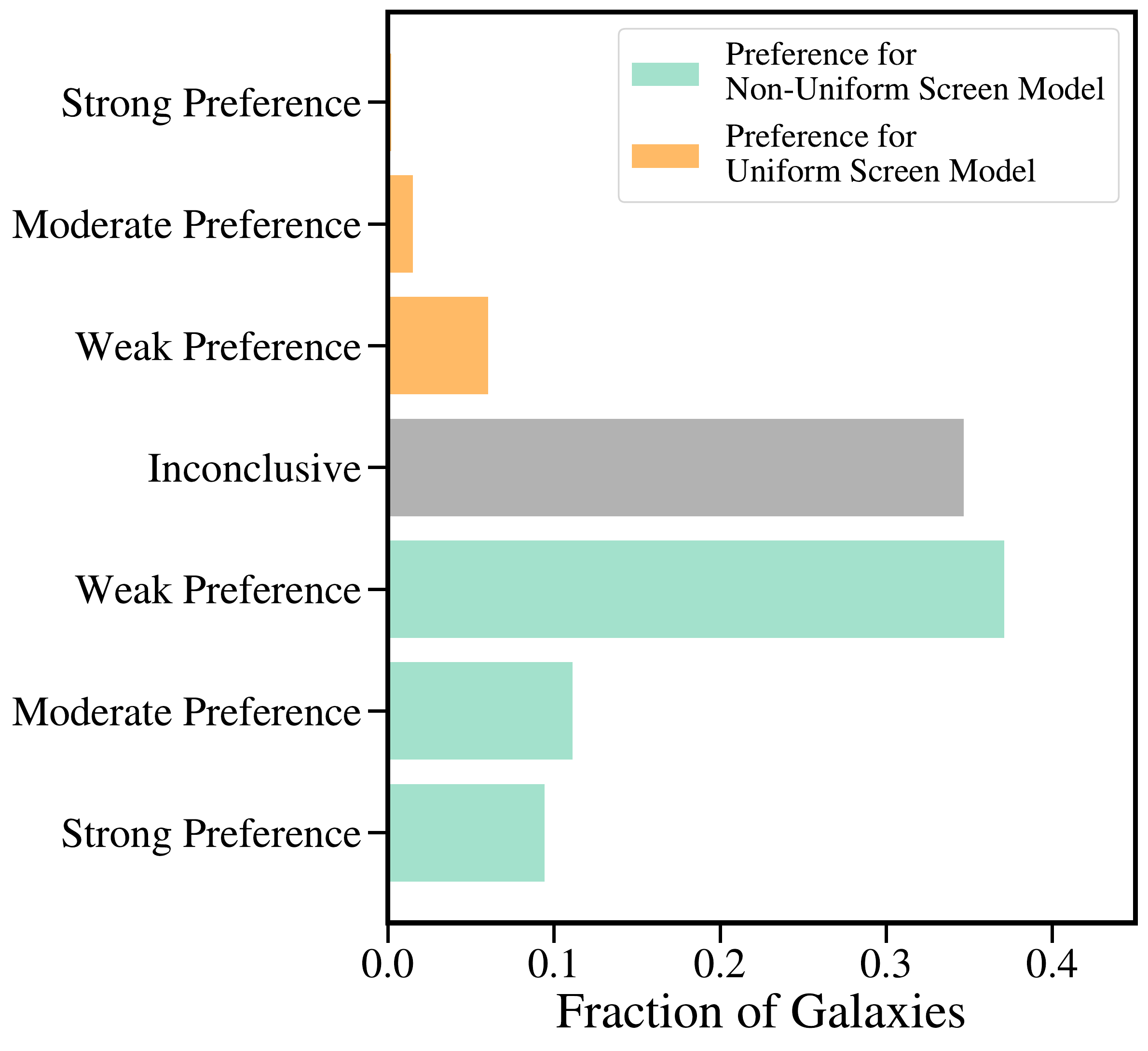}
    \caption{Plot showing the preference galaxies have for either variable slope model, quantified by the posterior odds of each model. The orange (green) bars show the fraction of galaxies that have a preference for the uniform (non-uniform) screen model, categorized by the strength of the evidence as outlined in \cite{kass_raftery_1995}. The grey bar represents the fraction of galaxies whose evidence does not prefer either model to a significant degree. A majority of galaxies have evidence that point to a preference for the non-uniform screen model.}
    \label{fig:evidence}
\end{figure}

\subsection{Constraining a More Flexible Attenuation Model}\label{sec:degen}

Though the results presented in \S \ref{sec:results} demonstrate an increased ability to constrain galaxy attenuation curves and physical properties with the non-uniform screen model, adding further complexity to an already multi-variate and physically complex model runs the risk of unconstrained results and over-fitting. This is compounded by our reliance on broadband photometry, which is relatively lower information than provided by, e.g., spectroscopy. The non-uniform model, or any dust attenuation model that is more complex than the typical variable slope model, could be more flexible than the data warrants and thus our results would be dominated by the priors and less by the likelihood. 

The use of {\sc dynesty}, and more generally Bayesian inference, means we can fully map the model parameter covariances and understand whether we suffer from over- or under-fitting from the use of a certain model. One way to measure this is by measuring the evidence in favor of one attenuation model over another. Following the methodology of \cite{salmon_2016}, we quantify this with the posterior odds, comparing the posterior marginalized over all model parameters of one dust attenuation model ($M$) to another given the available data ($D$):

\begin{equation}
    \mathrm{Posterior \: Odds} = \frac{p(M_1 | D)}{p(M_2 | D)}
\end{equation}
Given Bayes' theorem, $p(M _i| D)$ can be broken down into 

\begin{equation}
    \mathrm{Posterior \: Odds} = \frac{p(D | M_1)p(\theta M_1)}{p(D | M_2)p(\theta M_2)}
\end{equation}
where the first ratio is the Bayes factor and represents the integral of the posterior density over the parameter space for each model and the second ratio is the prior odds for model parameters $\theta$. The marginal likelihoods in the Bayes factor can be written as the integral over all parameter space for each model:

\begin{equation}
    p(D | M_i) = \int_\theta p(D | \theta, M_i)p(\theta | M_i) d\theta 
\end{equation}
which is also called the evidence. Thus, to calculate the posterior odds of two models, we take the ratio of their evidence and their prior probabilities, all of which are accessible with {\sc dynesty} nested sampling. 

Lastly, we use the criteria outlined in \cite{kass_raftery_1995} to denote the strength of the evidence towards one model or another. These criteria use the natural logarithm of the Bayes factor ($\ln(\mathrm{B}_{12})$) and place the posterior odds in categories of strong, moderate, or weak preference for one model over another; $\ln(\mathrm{B}_{12}) > 10$ denotes a strong preference for model 1 and $\ln(\mathrm{B}_{12}) < 2$ denotes a weak preference for model one. A preference does not necessarily point to one model being correct and the other incorrect. Rather, it describes the evidence against the opposing model, supporting the null hypothesis that the other model is correct. Strong, moderate, and weak preferences for a certain model are determined by the numerical value of the 

We calculate the posterior odds for the two variable slope attenuation curve models and evaluate the preference for one model over another for each galaxy using the criteria outlined in \cite{kass_raftery_1995}. In Figure \ref{fig:evidence}, we plot the fraction of galaxies which fall into each category (strong, moderate, weak preference) for either model. The fraction of galaxies whose evidence does not prefer either model are labeled inconclusive. A majority of galaxies ($61\%$) have Bayesian evidence that points to a preference for the non-uniform screen model while less than $9\%$ of galaxies have evidence that points towards the uniform screen model. Because Bayesian evidence is highly sensitive to the model prior space, we also fit SEDs with a more informative prior on f$_\mathrm{unobscured}$ based on the distribution in Figure \ref{fig:simba_props}. We find that the preference for the non-uniform screen model has fallen to $53\%$ of galaxies with $14\%$ preferring the uniform screen model. We interpret these results as 1) the more flexible model is preferred by a majority of galaxies with little preference (over inconclusive preference) for the less flexible model and 2) that while the prior space has moderate influence on the preference towards the non-uniform screen model there are sufficient constraints from the data for such a preference, alleviating some dangers of over-fitting by a more complex model than warranted by the data. 

\begin{figure}
    \centering
    \includegraphics[width=0.5\textwidth]{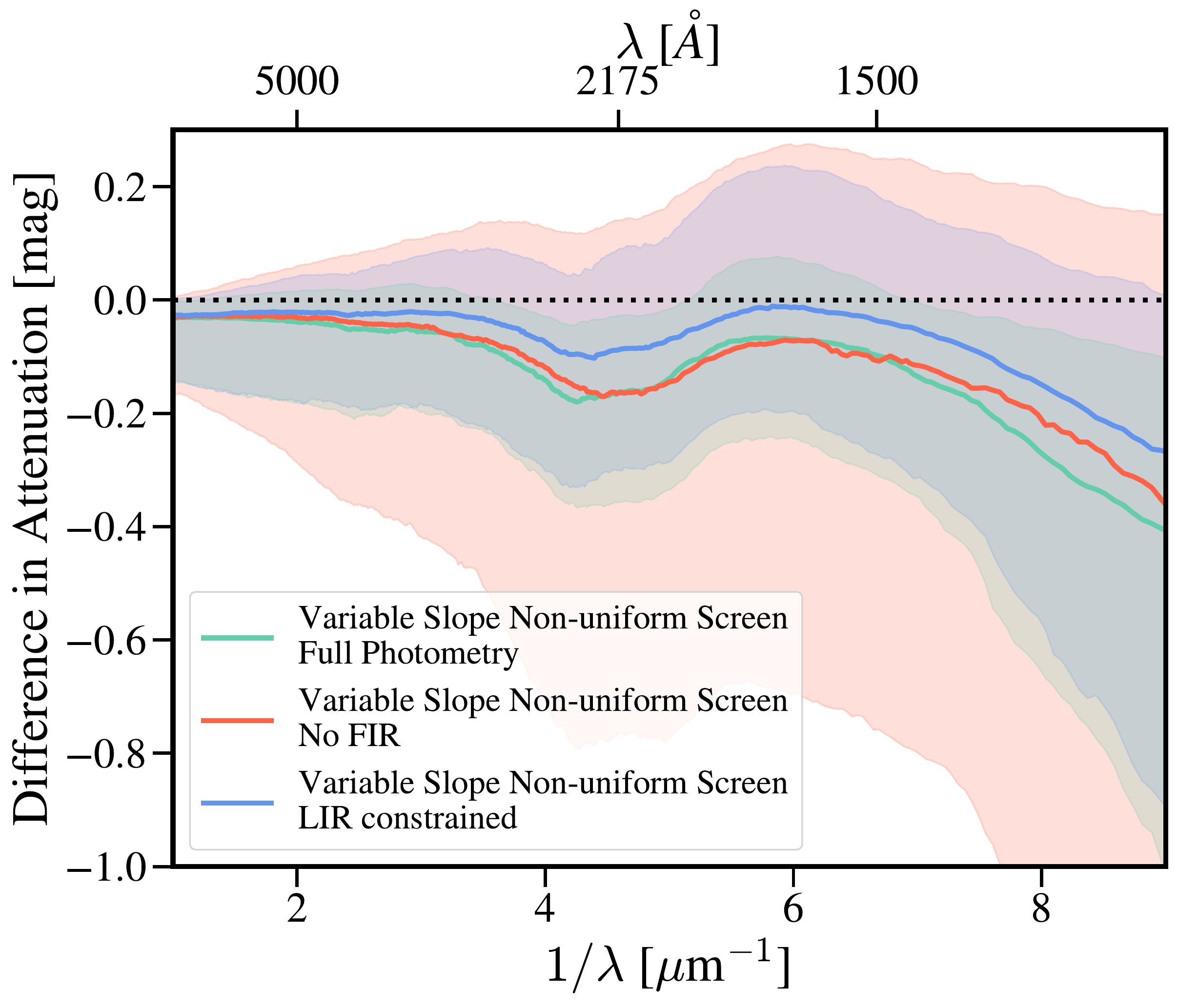}
    \caption{The distribution of attenuation curve offsets comparing the non-uniform screen model fit to different photometric data sets. The solid lines denote the median offset while the shaded regions represent the $16^{th}$ through $84^{th}$ percentiles.}
    \label{fig:fir}
\end{figure}

\subsection{Dependency on SED Coverage: FIR Photometry}\label{sec:fir_phot}

Finally, we consider the need for FIR constraints on our ability to accurately derive dust attenuation curves from SED fitting.  In specific,  far-infrared (FIR) data can break the reddening degeneracy between age and dust. Our results presented thus far have benefited from a fully sampled SED. We therefore investigate the efficacy of these techniques for observations that do not benefit from the availability of FIR data.  

To address this, we fit our mock SEDs without the \textit{Spitzer} and \textit{Herschel}  bands in the FIR and compare the inferred attenuation curves in Figure \ref{fig:fir}. We show the results from our fiducial model (non-uniform screen, fully sampled SED including FIR photometry) in green, while the  results from fitting with no FIR data are shown in red. We note that while the average bias between the true and inferred attenuation curves does not change, there is a significant increase in the scatter of this offset at fixed wavelength, implying a decreased ability to correctly infer the attenuation for an individual galaxy.

Because only the total infrared luminosity is needed to constrain the amount of dust attenuation in the UV and optical regime, one solution is to provide $L_{IR}$ as an input to the {\sc prospector} fit in a similar fashion as done by \cite{salim_2018} with the code {\sc CIGALE}. In Figure \ref{fig:fir}, the fits performed with $L_{IR}$ as a constraint are given in blue. The results show a significant decrease in scatter at a fixed wavelength: at $2300 \AA$ the distribution widths are $0.64$, $0.25$, and $0.21$ magnitudes for the no FIR, $L_{IR}$, and full photometry fits respectively. From this, we conclude that while a fully sampled FIR SED provides the best constraints for modeling the dust attenuation in a galaxy, approximating this with only the infrared luminosity sufficiently decreases the scatter in the inferred attenuation curves from fits with no FIR constraints.

\section{Conclusions}

Using galaxies from the {\sc simba} cosmological galaxy formation simulation, we have shown an increased ability to infer galaxy attenuation curves from SED modeling when employing a non-uniform screen model. This non-uniform screen, in contrast to the traditional uniform screen approximation, allows for a variable diffuse dust covering fraction such that a fraction of the stellar spectra is left unattenuated by dust. The non-uniform screen model introduces the flexibility necessary to recover the simulated dust attenuation curves and galaxy physical properties simultaneously. While this parameterization does not make explicit statements in the exact star-dust distribution, i.e. we cannot infer the exact coupling of the stars and dust nor do we model the covering fraction as a function of stellar age, it captures the aggregate effect of the star-dust geometry that deviates from the uniform screen limit. The key conclusions from our analysis are summarized below: 

\begin{enumerate}
    \item Compared to traditionally employed uniform screen models for dust attenuation, a non-uniform screen model significantly increases our ability to infer the shape of galaxy attenuation curves, as shown in Figure \ref{fig:attenuation_curves}.
    \item We also more robustly infer galaxy properties like SFR and metallicity with a non-uniform screen model, as demonstrated in Figures \ref{fig:sfr} and \ref{fig:a1500_sfr_Z}. This is directly tied to an improved ability to infer the magnitude of attenuation in the UV, as demonstrated in Figure \ref{fig:a1500_sfr_Z}. 
    \item Far-infrared observations may be necessary to place limits on the amount of stellar UV and optical light reprocessed by dust in the FIR. However, in the absence of a fully sampled FIR SED, an estimate for the total infrared luminosity can be used to constrain the model. We show in \S \ref{sec:fir_phot} and Figure \ref{fig:fir} that fitting for the total infrared luminosity, instead of individual photometry in the FIR, is sufficient to decrease confusion compared to a case where no FIR photometry is used.
    
\end{enumerate}

\section*{Acknowledgements} D.N. acknowledges support from NSF-1909153. C.C. acknowledges support from the Packard Foundation.

The {\sc simba} simulation used in this work was run on the ARCHER U.K. National Supercomputing Service, {\tt http://www.archer.ac.uk}.

\textit{Software}: {\sc simba} \citep{dave_simba}, {\sc caesar} \citep{caesar}, {\sc powderday} (\citealt{powderday}), {\sc hyperion} \citep{robitaille_2011_hyperion}, {\sc fsps} \citep{fsps_1, fsps_2}, {\sc python-fsps} \citep{dan_foreman_mackey_2014_12157}, {\sc prospector} \citep{2017zndo...1116491J, prosp}, {\sc numpy} \citep{5725236}, {\sc matplotlib} \citep{2018zndo...1482098C}, {\sc astropy} \citep{2020SciPy-NMeth}

\bibliography{bib}{}

\begin{thebibliography}{}
\expandafter\ifx\csname natexlab\endcsname\relax\def\natexlab#1{#1}\fi
\providecommand{\url}[1]{\href{#1}{#1}}
\providecommand{\dodoi}[1]{doi:~\href{http://doi.org/#1}{\nolinkurl{#1}}}
\providecommand{\doeprint}[1]{\href{http://ascl.net/#1}{\nolinkurl{http://ascl.net/#1}}}
\providecommand{\doarXiv}[1]{\href{https://arxiv.org/abs/#1}{\nolinkurl{https://arxiv.org/abs/#1}}}

\bibitem[{{Battisti} {et~al.}(2016){Battisti}, {Calzetti}, \&
  {Chary}}]{battisti_2016}
{Battisti}, A.~J., {Calzetti}, D., \& {Chary}, R.~R. 2016, \apj, 818, 13,
  \dodoi{10.3847/0004-637X/818/1/13}

\bibitem[{{Battisti} {et~al.}(2017{\natexlab{a}}){Battisti}, {Calzetti}, \&
  {Chary}}]{battisti_2017_1}
---. 2017{\natexlab{a}}, \apj, 840, 109, \dodoi{10.3847/1538-4357/aa6fb2}

\bibitem[{{Battisti} {et~al.}(2017{\natexlab{b}}){Battisti}, {Calzetti}, \&
  {Chary}}]{battisti_2017_2}
---. 2017{\natexlab{b}}, \apj, 851, 90, \dodoi{10.3847/1538-4357/aa9a43}

\bibitem[{{Betancourt} \& {Girolami}(2013)}]{Betancourt_2013_dirichlet}
{Betancourt}, M.~J., \& {Girolami}, M. 2013, arXiv e-prints, arXiv:1312.0906.
\newblock \doarXiv{1312.0906}

\bibitem[{{Bogdanoska} \& {Burgarella}(2020)}]{Bogdanoska_2020}
{Bogdanoska}, J., \& {Burgarella}, D. 2020, \mnras, 496, 5341,
  \dodoi{10.1093/mnras/staa1928}

\bibitem[{{Boquien} {et~al.}(2019){Boquien}, {Burgarella}, {Roehlly}, {Buat},
  {Ciesla}, {Corre}, {Inoue}, \& {Salas}}]{boquien_2019_cigale}
{Boquien}, M., {Burgarella}, D., {Roehlly}, Y., {et~al.} 2019, \aap, 622, A103,
  \dodoi{10.1051/0004-6361/201834156}

\bibitem[{{Burgarella} {et~al.}(2005){Burgarella}, {Buat}, \&
  {Iglesias-P{\'a}ramo}}]{cigale_1}
{Burgarella}, D., {Buat}, V., \& {Iglesias-P{\'a}ramo}, J. 2005, \mnras, 360,
  1413, \dodoi{10.1111/j.1365-2966.2005.09131.x}

\bibitem[{{Calzetti}(1997)}]{calzetti_1997}
{Calzetti}, D. 1997, \aj, 113, 162, \dodoi{10.1086/118242}

\bibitem[{{Calzetti}(2001)}]{calzetti_2001_geo}
---. 2001, \pasp, 113, 1449, \dodoi{10.1086/324269}

\bibitem[{{Calzetti} {et~al.}(2000){Calzetti}, {Armus}, {Bohlin}, {Kinney},
  {Koornneef}, \& {Storchi-Bergmann}}]{calzetti_2000}
{Calzetti}, D., {Armus}, L., {Bohlin}, R.~C., {et~al.} 2000, \apj, 533, 682,
  \dodoi{10.1086/308692}

\bibitem[{{Calzetti} {et~al.}(1994){Calzetti}, {Kinney}, \&
  {Storchi-Bergmann}}]{calzetti_1994}
{Calzetti}, D., {Kinney}, A.~L., \& {Storchi-Bergmann}, T. 1994, \apj, 429,
  582, \dodoi{10.1086/174346}

\bibitem[{{Cardelli} {et~al.}(1989){Cardelli}, {Clayton}, \&
  {Mathis}}]{ccm_1989_mw_ext}
{Cardelli}, J.~A., {Clayton}, G.~C., \& {Mathis}, J.~S. 1989, \apj, 345, 245,
  \dodoi{10.1086/167900}

\bibitem[{{Casey} {et~al.}(2014){Casey}, {Scoville}, {Sanders}, {Lee},
  {Cooray}, {Finkelstein}, {Capak}, {Conley}, {De Zotti}, {Farrah}, {Fu}, {Le
  Floc'h}, {Ilbert}, {Ivison}, \& {Takeuchi}}]{casey_2014}
{Casey}, C.~M., {Scoville}, N.~Z., {Sanders}, D.~B., {et~al.} 2014, \apj, 796,
  95, \dodoi{10.1088/0004-637X/796/2/95}

\bibitem[{{Casey} {et~al.}(2018){Casey}, {Zavala}, {Spilker}, {da Cunha},
  {Hodge}, {Hung}, {Staguhn}, {Finkelstein}, \& {Drew}}]{casey_2018}
{Casey}, C.~M., {Zavala}, J.~A., {Spilker}, J., {et~al.} 2018, \apj, 862, 77,
  \dodoi{10.3847/1538-4357/aac82d}

\bibitem[{{Caswell} {et~al.}(2018){Caswell}, {Droettboom}, {Hunter}, {Firing},
  {Lee}, {Stansby}, {Sales de Andrade}, {Hedegaard Nielsen}, {Klymak},
  {Varoquaux}, {Root}, {Elson}, {Dale}, {May}, {Lee}, {Sepp{\"a}nen},
  {Hoffmann}, {McDougall}, {Straw}, {Hobson}, {cgohlke}, {Yu}, {Ma}, {Vincent},
  {Silvester}, {Moad}, {Katins}, {Kniazev}, {Ariza}, \&
  {W{\"u}rtz}}]{2018zndo...1482098C}
{Caswell}, T.~A., {Droettboom}, M., {Hunter}, J., {et~al.} 2018,
  {Matplotlib/Matplotlib V3.0.1}, v3.0.1,  Zenodo,
  \dodoi{10.5281/zenodo.1482098}

\bibitem[{{Charlot} \& {Fall}(2000)}]{charlot_fall_2000}
{Charlot}, S., \& {Fall}, S.~M. 2000, \apj, 539, 718, \dodoi{10.1086/309250}

\bibitem[{{Choi} {et~al.}(2016){Choi}, {Dotter}, {Conroy}, {Cantiello},
  {Paxton}, \& {Johnson}}]{choi_2016_mist}
{Choi}, J., {Dotter}, A., {Conroy}, C., {et~al.} 2016, \apj, 823, 102,
  \dodoi{10.3847/0004-637X/823/2/102}

\bibitem[{{Cleri} {et~al.}(2020){Cleri}, {Trump}, {Backhaus}, {Momcheva},
  {Papovich}, {Simons}, {Weiner}, {Estrada-Carpenter}, {Finkelstein},
  {Giavalisco}, {Ji}, {Jung}, {Matharu}, {Martinez}, \&
  {Sturm}}]{cleri_2020_pabeta}
{Cleri}, N.~J., {Trump}, J.~R., {Backhaus}, B.~E., {et~al.} 2020, arXiv
  e-prints, arXiv:2009.00617.
\newblock \doarXiv{2009.00617}

\bibitem[{{Conroy}(2013)}]{conroy_2013_araa}
{Conroy}, C. 2013, \araa, 51, 393, \dodoi{10.1146/annurev-astro-082812-141017}

\bibitem[{Conroy \& Gunn(2010)}]{fsps_2}
Conroy, C., \& Gunn, J.~E. 2010, The Astrophysical Journal, 712, 833–857,
  \dodoi{10.1088/0004-637x/712/2/833}

\bibitem[{Conroy {et~al.}(2009)Conroy, Gunn, \& White}]{fsps_1}
Conroy, C., Gunn, J.~E., \& White, M. 2009, The Astrophysical Journal, 699,
  486–506, \dodoi{10.1088/0004-637x/699/1/486}

\bibitem[{{Corre} {et~al.}(2018){Corre}, {Buat}, {Basa}, {Boissier}, {Japelj},
  {Palmerio}, {Salvaterra}, {Vergani}, \& {Zafar}}]{corre_2018}
{Corre}, D., {Buat}, V., {Basa}, S., {et~al.} 2018, \aap, 617, A141,
  \dodoi{10.1051/0004-6361/201832926}

\bibitem[{{Dav{\'e}} {et~al.}(2019){Dav{\'e}}, {Angl{\'e}s-Alc{\'a}zar},
  {Narayanan}, {Li}, {Rafieferantsoa}, \& {Appleby}}]{dave_simba}
{Dav{\'e}}, R., {Angl{\'e}s-Alc{\'a}zar}, D., {Narayanan}, D., {et~al.} 2019,
  \mnras, 486, 2827, \dodoi{10.1093/mnras/stz937}

\bibitem[{{Dotter}(2016)}]{dotter_2016_mist}
{Dotter}, A. 2016, \apjs, 222, 8, \dodoi{10.3847/0067-0049/222/1/8}

\bibitem[{{Draine}(2003)}]{draine_03}
{Draine}, B.~T. 2003, \araa, 41, 241,
  \dodoi{10.1146/annurev.astro.41.011802.094840}

\bibitem[{Draine \& Li(2007)}]{draine_infrared_2007}
Draine, B.~T., \& Li, A. 2007, The Astrophysical Journal, 657, 810,
  \dodoi{10.1086/511055}

\bibitem[{Foreman-Mackey {et~al.}(2014)Foreman-Mackey, Sick, \&
  Johnson}]{dan_foreman_mackey_2014_12157}
Foreman-Mackey, D., Sick, J., \& Johnson, B. 2014, python-fsps: Python bindings
  to FSPS (v0.1.1), v0.1.1,  Zenodo, \dodoi{10.5281/zenodo.12157}

\bibitem[{Gallazzi {et~al.}(2005)Gallazzi, Charlot, Brinchmann, White, \&
  Tremonti}]{gallazzi_ages_2005}
Gallazzi, A., Charlot, S., Brinchmann, J., White, S. D.~M., \& Tremonti, C.~A.
  2005, Monthly Notices of the Royal Astronomical Society, 362, 41,
  \dodoi{10.1111/j.1365-2966.2005.09321.x}

\bibitem[{{Hagen} {et~al.}(2017){Hagen}, {Siegel}, {Hoversten}, {Gronwall},
  {Immler}, \& {Hagen}}]{hagen_2017}
{Hagen}, L. M.~Z., {Siegel}, M.~H., {Hoversten}, E.~A., {et~al.} 2017, \mnras,
  466, 4540, \dodoi{10.1093/mnras/stw2954}

\bibitem[{{Hayward} \& {Smith}(2015)}]{hayward_smith_2015}
{Hayward}, C.~C., \& {Smith}, D. J.~B. 2015, \mnras, 446, 1512,
  \dodoi{10.1093/mnras/stu2195}

\bibitem[{{Hopkins}(2015)}]{hopkins_2015_gizmo}
{Hopkins}, P.~F. 2015, \mnras, 450, 53, \dodoi{10.1093/mnras/stv195}

\bibitem[{{Johnson} \& {Leja}(2017)}]{2017zndo...1116491J}
{Johnson}, B., \& {Leja}, J. 2017, {Bd-J/Prospector: Initial Release}, v0.1,
  Zenodo, \dodoi{10.5281/zenodo.1116491}

\bibitem[{{Johnson} {et~al.}(2021){Johnson}, {Leja}, {Conroy}, \&
  {Speagle}}]{prosp}
{Johnson}, B.~D., {Leja}, J., {Conroy}, C., \& {Speagle}, J.~S. 2021, \apjs,
  254, 22, \dodoi{10.3847/1538-4365/abef67}

\bibitem[{Kass \& Raftery(1995)}]{kass_raftery_1995}
Kass, R.~E., \& Raftery, A.~E. 1995, Journal of the American Statistical
  Association, 90, 773, \dodoi{10.1080/01621459.1995.10476572}

\bibitem[{Kriek \& Conroy(2013)}]{kc_law}
Kriek, M., \& Conroy, C. 2013, The Astrophysical Journal, 775, L16,
  \dodoi{10.1088/2041-8205/775/1/L16}

\bibitem[{{Kriek} {et~al.}(2018){Kriek}, {van Dokkum}, {Labb{\'e}}, {Franx},
  {Illingworth}, {Marchesini}, {Quadri}, {Aird}, {Coil}, \&
  {Georgakakis}}]{fast_software}
{Kriek}, M., {van Dokkum}, P.~G., {Labb{\'e}}, I., {et~al.} 2018, {FAST:
  Fitting and Assessment of Synthetic Templates}.
\newblock \doeprint{1803.008}

\bibitem[{{Kroupa}(2002)}]{kroupa_2002_imf}
{Kroupa}, P. 2002, Science, 295, 82, \dodoi{10.1126/science.1067524}

\bibitem[{{Leja} {et~al.}(2019){Leja}, {Carnall}, {Johnson}, {Conroy}, \&
  {Speagle}}]{leja_2019_nonpara}
{Leja}, J., {Carnall}, A.~C., {Johnson}, B.~D., {Conroy}, C., \& {Speagle},
  J.~S. 2019, \apj, 876, 3, \dodoi{10.3847/1538-4357/ab133c}

\bibitem[{Leja {et~al.}(2017)Leja, Johnson, Conroy, van Dokkum, \&
  Byler}]{leja_deriving_2017}
Leja, J., Johnson, B.~D., Conroy, C., van Dokkum, P.~G., \& Byler, N. 2017, The
  Astrophysical Journal, 837, 170, \dodoi{10.3847/1538-4357/aa5ffe}

\bibitem[{{Leslie} {et~al.}(2018){Leslie}, {Sargent}, {Schinnerer}, {Groves},
  {van der Wel}, {Zamorani}, {Fudamoto}, {Lang}, \&
  {Smol{\v{c}}i{\'c}}}]{leslie_2018_clumpiness}
{Leslie}, S.~K., {Sargent}, M.~T., {Schinnerer}, E., {et~al.} 2018, \aap, 615,
  A7, \dodoi{10.1051/0004-6361/201732255}

\bibitem[{{Li} {et~al.}(2019){Li}, {Narayanan}, \& {Dav{\'e}}}]{li_2019_dust}
{Li}, Q., {Narayanan}, D., \& {Dav{\'e}}, R. 2019, \mnras, 490, 1425,
  \dodoi{10.1093/mnras/stz2684}

\bibitem[{{Lower} {et~al.}(2020){Lower}, {Narayanan}, {Leja}, {Johnson},
  {Conroy}, \& {Dav{\'e}}}]{lower_stellar_mass}
{Lower}, S., {Narayanan}, D., {Leja}, J., {et~al.} 2020, arXiv e-prints,
  arXiv:2006.03599.
\newblock \doarXiv{2006.03599}

\bibitem[{{Lucy}(1999)}]{lucy_rt}
{Lucy}, L.~B. 1999, \aap, 344, 282

\bibitem[{{Madau} \& {Dickinson}(2014)}]{madau_dickinson_2014}
{Madau}, P., \& {Dickinson}, M. 2014, \araa, 52, 415,
  \dodoi{10.1146/annurev-astro-081811-125615}

\bibitem[{{Narayanan} {et~al.}(2018){Narayanan}, {Conroy}, {Dav{\'e}},
  {Johnson}, \& {Popping}}]{narayanan_2018_atten}
{Narayanan}, D., {Conroy}, C., {Dav{\'e}}, R., {Johnson}, B.~D., \& {Popping},
  G. 2018, \apj, 869, 70, \dodoi{10.3847/1538-4357/aaed25}

\bibitem[{{Narayanan} {et~al.}(2021){Narayanan}, {Turk}, {Robitaille}, {Kelly},
  {McClellan}, {Sharma}, {Garg}, {Abruzzo}, {Choi}, {Conroy}, {Johnson},
  {Kimock}, {Li}, {Lovell}, {Lower}, {Privon}, {Roberts}, {Sethuram}, {Snyder},
  {Thompson}, \& {Wise}}]{powderday}
{Narayanan}, D., {Turk}, M.~J., {Robitaille}, T., {et~al.} 2021, \apjs, 252,
  12, \dodoi{10.3847/1538-4365/abc487}

\bibitem[{{Natale} {et~al.}(2015){Natale}, {Popescu}, {Tuffs}, {Debattista},
  {Fischera}, \& {Grootes}}]{natale_2015}
{Natale}, G., {Popescu}, C.~C., {Tuffs}, R.~J., {et~al.} 2015, \mnras, 449,
  243, \dodoi{10.1093/mnras/stv286}

\bibitem[{{Noll} {et~al.}(2009{\natexlab{a}}){Noll}, {Burgarella},
  {Giovannoli}, {Buat}, {Marcillac}, \& {Mu{\~n}oz-Mateos}}]{noll_2009}
{Noll}, S., {Burgarella}, D., {Giovannoli}, E., {et~al.} 2009{\natexlab{a}},
  \aap, 507, 1793, \dodoi{10.1051/0004-6361/200912497}

\bibitem[{{Noll} {et~al.}(2009{\natexlab{b}}){Noll}, {Burgarella},
  {Giovannoli}, {Buat}, {Marcillac}, \& {Mu{\~n}oz-Mateos}}]{cigale_2}
---. 2009{\natexlab{b}}, \aap, 507, 1793, \dodoi{10.1051/0004-6361/200912497}

\bibitem[{{Noll} {et~al.}(2007){Noll}, {Pierini}, {Pannella}, \&
  {Savaglio}}]{noll_2007}
{Noll}, S., {Pierini}, D., {Pannella}, M., \& {Savaglio}, S. 2007, \aap, 472,
  455, \dodoi{10.1051/0004-6361:20077067}

\bibitem[{{Pei}(1992)}]{pei_1992_smc}
{Pei}, Y.~C. 1992, \apj, 395, 130, \dodoi{10.1086/171637}

\bibitem[{{Popping} {et~al.}(2017){Popping}, {Puglisi}, \&
  {Norman}}]{popping_2017}
{Popping}, G., {Puglisi}, A., \& {Norman}, C.~A. 2017, \mnras, 472, 2315,
  \dodoi{10.1093/mnras/stx2202}

\bibitem[{{Qin} {et~al.}(2022){Qin}, {Zheng}, {Fang}, {Pan}, {Wuyts}, {Shi},
  {Peng}, {Gonzalez}, {Bian}, {Huang}, {Gu}, {Liu}, {Tan}, {Shi}, {Ren},
  {Zhang}, {Qiao}, {Wen}, \& {Liu}}]{2022MNRAS.tmp..190Q}
{Qin}, J., {Zheng}, X.~Z., {Fang}, M., {et~al.} 2022, \mnras,
  \dodoi{10.1093/mnras/stac132}

\bibitem[{{Reddy} {et~al.}(2015){Reddy}, {Kriek}, {Shapley}, {Freeman},
  {Siana}, {Coil}, {Mobasher}, {Price}, {Sanders}, \& {Shivaei}}]{reddy_2015}
{Reddy}, N.~A., {Kriek}, M., {Shapley}, A.~E., {et~al.} 2015, \apj, 806, 259,
  \dodoi{10.1088/0004-637X/806/2/259}

\bibitem[{{Reddy} {et~al.}(2018){Reddy}, {Oesch}, {Bouwens}, {Montes},
  {Illingworth}, {Steidel}, {van Dokkum}, {Atek}, {Carollo}, {Cibinel},
  {Holden}, {Labb{\'e}}, {Magee}, {Morselli}, {Nelson}, \&
  {Wilkins}}]{reddy_2018}
{Reddy}, N.~A., {Oesch}, P.~A., {Bouwens}, R.~J., {et~al.} 2018, \apj, 853, 56,
  \dodoi{10.3847/1538-4357/aaa3e7}

\bibitem[{{Reddy} {et~al.}(2020){Reddy}, {Shapley}, {Kriek}, {Steidel},
  {Shivaei}, {Sanders}, {Mobasher}, {Coil}, {Siana}, {Freeman}, {Azadi},
  {Fetherolf}, {Leung}, {Price}, \& {Zick}}]{reddy_2020_neb}
{Reddy}, N.~A., {Shapley}, A.~E., {Kriek}, M., {et~al.} 2020, \apj, 902, 123,
  \dodoi{10.3847/1538-4357/abb674}

\bibitem[{{Robitaille}(2011)}]{robitaille_2011_hyperion}
{Robitaille}, T.~P. 2011, \aap, 536, A79, \dodoi{10.1051/0004-6361/201117150}

\bibitem[{{Robitaille} {et~al.}(2012){Robitaille}, {Churchwell}, {Benjamin},
  {Whitney}, {Wood}, {Babler}, \& {Meade}}]{robitaille_pahs}
{Robitaille}, T.~P., {Churchwell}, E., {Benjamin}, R.~A., {et~al.} 2012, \aap,
  545, A39, \dodoi{10.1051/0004-6361/201219073}

\bibitem[{{Salim} {et~al.}(2018){Salim}, {Boquien}, \& {Lee}}]{salim_2018}
{Salim}, S., {Boquien}, M., \& {Lee}, J.~C. 2018, \apj, 859, 11,
  \dodoi{10.3847/1538-4357/aabf3c}

\bibitem[{{Salim} \& {Narayanan}(2020)}]{attenuation_araa}
{Salim}, S., \& {Narayanan}, D. 2020, \araa, 58, 529,
  \dodoi{10.1146/annurev-astro-032620-021933}

\bibitem[{{Salim} {et~al.}(2016){Salim}, {Lee}, {Janowiecki}, {da Cunha},
  {Dickinson}, {Boquien}, {Burgarella}, {Salzer}, \& {Charlot}}]{salim_2016}
{Salim}, S., {Lee}, J.~C., {Janowiecki}, S., {et~al.} 2016, \apjs, 227, 2,
  \dodoi{10.3847/0067-0049/227/1/2}

\bibitem[{{Salmon} {et~al.}(2016){Salmon}, {Papovich}, {Long}, {Willner},
  {Finkelstein}, {Ferguson}, {Dickinson}, {Duncan}, {Faber}, {Hathi},
  {Koekemoer}, {Kurczynski}, {Newman}, {Pacifici}, {P{\'e}rez-Gonz{\'a}lez}, \&
  {Pforr}}]{salmon_2016}
{Salmon}, B., {Papovich}, C., {Long}, J., {et~al.} 2016, \apj, 827, 20,
  \dodoi{10.3847/0004-637X/827/1/20}

\bibitem[{{Seon} \& {Draine}(2016)}]{seon_draine_2016}
{Seon}, K.-I., \& {Draine}, B.~T. 2016, \apj, 833, 201,
  \dodoi{10.3847/1538-4357/833/2/201}

\bibitem[{{Sharma} {et~al.}(2021){Sharma}, {Choi}, {Somerville}, {Snyder},
  {Kocevski}, {Hirschmann}, {Moster}, {Naab}, {Narayanan}, {Ostriker}, \&
  {Rosario}}]{sharma_2021}
{Sharma}, R.~S., {Choi}, E., {Somerville}, R.~S., {et~al.} 2021, arXiv
  e-prints, arXiv:2101.01729.
\newblock \doarXiv{2101.01729}

\bibitem[{{Speagle}(2020)}]{dynesty}
{Speagle}, J.~S. 2020, \mnras, 493, 3132, \dodoi{10.1093/mnras/staa278}

\bibitem[{{Thompson}(2014)}]{caesar}
{Thompson}, R. 2014, {pyGadgetReader: GADGET snapshot reader for python}.
\newblock \doeprint{1411.001}

\bibitem[{{Trayford} {et~al.}(2020){Trayford}, {Lagos}, {Robotham}, \&
  {Obreschkow}}]{trayford_2020}
{Trayford}, J.~W., {Lagos}, C. d.~P., {Robotham}, A. S.~G., \& {Obreschkow}, D.
  2020, \mnras, 491, 3937, \dodoi{10.1093/mnras/stz3234}

\bibitem[{{Trayford} {et~al.}(2017){Trayford}, {Camps}, {Theuns}, {Baes},
  {Bower}, {Crain}, {Gunawardhana}, {Schaller}, {Schaye}, \&
  {Frenk}}]{trayford_2017}
{Trayford}, J.~W., {Camps}, P., {Theuns}, T., {et~al.} 2017, \mnras, 470, 771,
  \dodoi{10.1093/mnras/stx1051}

\bibitem[{{Tress} {et~al.}(2018){Tress}, {M{\'a}rmol-Queralt{\'o}}, {Ferreras},
  {P{\'e}rez-Gonz{\'a}lez}, {Barro}, {Pampliega}, {Cava},
  {Dom{\'\i}nguez-S{\'a}nchez}, {Eliche-Moral}, {Espino-Briones}, {Esquej},
  {Hern{\'a}n-Caballero}, {Rodighiero}, \& {Rodriguez-Mu{\~n}oz}}]{tress_2018}
{Tress}, M., {M{\'a}rmol-Queralt{\'o}}, E., {Ferreras}, I., {et~al.} 2018,
  \mnras, 475, 2363, \dodoi{10.1093/mnras/stx3334}

\bibitem[{{Tuffs} {et~al.}(2004){Tuffs}, {Popescu}, {V{\"o}lk}, {Kylafis}, \&
  {Dopita}}]{tuffs_2004_clumpiness}
{Tuffs}, R.~J., {Popescu}, C.~C., {V{\"o}lk}, H.~J., {Kylafis}, N.~D., \&
  {Dopita}, M.~A. 2004, \aap, 419, 821, \dodoi{10.1051/0004-6361:20035689}

\bibitem[{{Turk} {et~al.}(2011){Turk}, {Smith}, {Oishi}, {Skory}, {Skillman},
  {Abel}, \& {Norman}}]{yt}
{Turk}, M.~J., {Smith}, B.~D., {Oishi}, J.~S., {et~al.} 2011, \apjs, 192, 9,
  \dodoi{10.1088/0067-0049/192/1/9}

\bibitem[{{Valencic} {et~al.}(2004){Valencic}, {Clayton}, \&
  {Gordon}}]{valencic_2004_MW_bump_strength}
{Valencic}, L.~A., {Clayton}, G.~C., \& {Gordon}, K.~D. 2004, \apj, 616, 912,
  \dodoi{10.1086/424922}

\bibitem[{van~der Giessen {et~al.}(2022)van~der Giessen, Leslie, Groves, Hodge,
  Popescu, Sargent, Schinnerer, \& Tuffs}]{vandergiessen_2022_dust_clumpiness}
van~der Giessen, S.~A., Leslie, S.~K., Groves, B., {et~al.} 2022.
\newblock \doarXiv{2201.10568}

\bibitem[{{van der Walt} {et~al.}(2011){van der Walt}, {Colbert}, \&
  {Varoquaux}}]{5725236}
{van der Walt}, S., {Colbert}, S.~C., \& {Varoquaux}, G. 2011, Computing in
  Science Engineering, 13, 22

\bibitem[{{Virtanen} {et~al.}(2020){Virtanen}, {Gommers}, {Oliphant},
  {Haberland}, {Reddy}, {Cournapeau}, {Burovski}, {Peterson}, {Weckesser},
  {Bright}, {van der Walt}, {Brett}, {Wilson}, {Jarrod Millman}, {Mayorov},
  {Nelson}, {Jones}, {Kern}, {Larson}, {Carey}, {Polat}, {Feng}, {Moore}, {Vand
  erPlas}, {Laxalde}, {Perktold}, {Cimrman}, {Henriksen}, {Quintero}, {Harris},
  {Archibald}, {Ribeiro}, {Pedregosa}, {van Mulbregt}, \&
  {Contributors}}]{2020SciPy-NMeth}
{Virtanen}, P., {Gommers}, R., {Oliphant}, T.~E., {et~al.} 2020, Nature
  Methods, 17, 261, \dodoi{https://doi.org/10.1038/s41592-019-0686-2}

\bibitem[{{Walcher} {et~al.}(2011){Walcher}, {Groves}, {Budav{\'a}ri}, \&
  {Dale}}]{walcher_2011_araa}
{Walcher}, J., {Groves}, B., {Budav{\'a}ri}, T., \& {Dale}, D. 2011, \apss,
  331, 1, \dodoi{10.1007/s10509-010-0458-z}

\bibitem[{{Weingartner} \& {Draine}(2001)}]{weingartner_draine}
{Weingartner}, J.~C., \& {Draine}, B.~T. 2001, \apj, 548, 296,
  \dodoi{10.1086/318651}

\bibitem[{{Witt} \& {Gordon}(1996)}]{witt_gordon_1996}
{Witt}, A.~N., \& {Gordon}, K.~D. 1996, \apj, 463, 681, \dodoi{10.1086/177282}

\bibitem[{{Witt} \& {Gordon}(2000)}]{witt_gordon_2000}
---. 2000, \apj, 528, 799, \dodoi{10.1086/308197}

\bibitem[{{Zavala} {et~al.}(2021){Zavala}, {Casey}, {Manning}, {Aravena},
  {Bethermin}, {Caputi}, {Clements}, {da Cunha}, {Drew}, {Finkelstein},
  {Fujimoto}, {Hayward}, {Hodge}, {Kartaltepe}, {Knudsen}, {Koekemoer}, {Long},
  {Magdis}, {Man}, {Popping}, {Sanders}, {Scoville}, {Sheth}, {Staguhn},
  {Toft}, {Treister}, {Vieira}, \& {Yun}}]{zavala_2021}
{Zavala}, J.~A., {Casey}, C.~M., {Manning}, S.~M., {et~al.} 2021, arXiv
  e-prints, arXiv:2101.04734.
\newblock \doarXiv{2101.04734}

\bibitem[{{Zuckerman} {et~al.}(2021){Zuckerman}, {Belli}, {Leja}, \&
  {Tacchella}}]{zuckerman_2021_inclination}
{Zuckerman}, L.~D., {Belli}, S., {Leja}, J., \& {Tacchella}, S. 2021, \apjl,
  922, L32, \dodoi{10.3847/2041-8213/ac3831}

\end{thebibliography}
\bibliographystyle{aasjournal}

\end{document}